\documentclass[a4paper, 11pt]{article}
\usepackage{amsmath,amsthm,amssymb,mathrsfs,mathtools,graphicx,color,yhmath}
\usepackage[titletoc,title]{appendix}
\usepackage[numbers ,sort&compress]{natbib}
\newcommand{\nc}{\newcommand}
\nc{\rnc}{\renewcommand}
\nc{\nn}{\nonumber}

\nc{\del}{{\partial}}
\rnc{\Im}{{\mathrm{Im}\,}}
\rnc{\Re}{{\mathrm{Re}\,}}

\nc{\bra}{\langle}
\nc{\ket}{\rangle}

\nc{\A}{\mathrm{A}}
\nc{\B}{\mathrm{B}}
\nc{\C}{\mathrm{C}}
\nc{\D}{\mathrm{D}}
\nc{\tcr}{\textcolor{red}}
\nc{\tcb}{\textcolor{blue}}

\theoremstyle{definition}

\numberwithin{equation}{section}

\begin{document}

\title{Two-level Quantum Walkers on Directed Graphs II: \\
An Application to qRAM}

\author{
Ryo Asaka\thanks{E-mail: hello.ryoasaka@gmail.com}, \,
Kazumitsu Sakai\thanks{E-mail: k.sakai@rs.tus.ac.jp} \, and
Ryoko Yahagi \thanks{E-mail: yahagi@rs.tus.ac.jp}
\\\\
\textit{Department of Physics,
Tokyo University of Science,}\\
 \textit{Kagurazaka 1-3, Shinjuku-ku, Tokyo 162-8601, Japan} \\
\\\\
\\
}

\date{April 19, 2022}

\maketitle

\begin{abstract}
This is the second paper in a series of two. Using a multi-particle 
continuous-time quantum walk with two internal states, which has been
formulated in the first paper (arXiv:2112.08119), we physically 
implement a quantum random access memory (qRAM). 
Data with address information are dual-rail encoded into 
quantum walkers.  The walkers pass through perfect binary 
trees to access the designated memory cells and copy the data stored 
in the cells. A roundabout gate allocated 
at each node serves as a router to move the walker from the parent node 
to one of two child nodes, depending on the internal state of the 
walker. In this process, the address information is sequentially 
encoded into the internal states so that the walkers are adequately 
delivered to the target cells.  
The present qRAM, which processes $2^n$ $m$-qubit data, is implemented 
in a quantum circuit of depth $O(n\log(n+m))$ and requires $O(n+m)$ qubit 
resources. 
This is more efficient than the conventional bucket-brigade qRAM that 
requires $O(n^2+nm)$ steps and $O(2^{n}+m)$ qubit resources for processing. 
Moreover, since the walkers are not entangled with any device on the 
binary trees, the cost of maintaining coherence can be reduced.
Notably, by simply passing quantum walkers through binary trees, 
data can be automatically extracted in a quantum superposition state. 
In other words,  any time-dependent control is not required. 
\end{abstract}
%%%%%%%%%%%%%%%%%%%%%%%%%%%%%%%%%%%%%%%%%%%%%%%%%%%%%%%%%%%%
\section{Introduction}\label{introduction}
%%%%%%%%%%%%%%%%%%%%%%%%%%%%%%%%%%%%%%%%%%%%%%%%%%%%%%%%%%%%
This is the second paper in a series of two in which we consider
a multi-particle continuous-time quantum walk with two internal states.
In the present paper, we propose
a physical implementation of a quantum random access memory (qRAM),
using some devices developed in the first paper \cite{asaka2021two}, 
in which an architecture of universal quantum computation using the 
quantum walk has been provided.

A number of quantum algorithms exploiting quantum 
mechanical effects have been proposed to achieve significant speedups 
over their classical analogs \cite{nielsen2002quantum}. Algorithms for 
quantum phase estimation \cite{shor1994algorithms,kitaev1995quantum}, 
quantum amplitude amplification \cite{brassard1997exact,brassard2002quantum,
grover1996fast,grover1998quantum}, and quantum Hamiltonian simulation 
\cite{feynman2018simulating,lloyd1996universal,berry2007efficient,
berry2012black,berry2015hamiltonian,berry2015simulating,
low2017optimal,childs2018toward,low2019hamiltonian,bauer2020quantum} 
are the most notable, and are used as subroutines in, for example, Shor's 
algorithm for factorizing large integers \cite{shor1994algorithms} and
Grover's algorithm for searching unsorted databases \cite{grover1996fast}.  
However, one should be careful about claiming that quantum algorithms are 
superior to classical counterparts in some cases. As an example, let us take 
the search problem of finding a particular item in an unstructured set 
consisting of $N$ items. Grover's algorithm incorporates the process of
accessing and querying the database as an oracle (a black box that answers 
yes or no) and completes the search with only $O(\sqrt{N})$ oracle 
queries, achieving a quadratic speedup over classical exhaustive search. 
In practice, however, the oracle subroutines, i.e., converting data into 
a quantum superposition state, accessing and reading them, maybe a 
cumbersome overhead that offsets the quantum speedup \cite{viamontes2005quantum}.
Namely, reducing a cost to the oracle is crucial for 
applications of quantum computation to search problems, Hamiltonian 
simulations and machine learning for big data \cite{lloyd2013quantum,
rebentrost2014quantum,wittek2014quantum,prakash2014quantum,
schuld2015introduction,adcock2015advances,biamonte2017quantum,
schuld2017implementing,dunjko2018machine,ciliberto2018quantum,
kerenidis2018q,gao2018quantum,schuld2019quantum,bang2019optimal,
ramezani2020machine,zhang2020recent,
zhou2021blind,mishra2021quantum,jiang2021machine}.

A quantum random access memory (qRAM) was introduced as a quantum counterpart 
of a  RAM, 
promising to efficiently access data and convert them into superposition 
states \cite{giovannetti2008quantum,giovannetti2008architectures}.
Conceptually, a qRAM is a quantum device comprising the following three principal schemes: (i)
a routing scheme to access the specified memory cells whose addresses
are given by an $n$-qubit superposition state 
\begin{equation}
\sum_a |a\ket_A=\sum_{\{a_j\}} | a_{n-1}\cdots a_0\ket_A
\in(\mathbb{C}^2)^{\otimes n},
\quad a\in\mathbb{Z}_{\ge 0},
\,\, a_{j}\in\{0,1\}\,\, (0\le j \le n-1),
\label{address}
\end{equation}
(ii) a querying scheme to read the classical information
$x^{(a)}\in \mathbb{Z}_{\ge 0}$ stored in the $a$th cell\footnote{A qRAM 
can also process the quantum information where
$|x^{(a)}\ket$ consists of a superposition of states. See Sec.~5. 
For the moment, however, we restrict ourselves to the classical case for convenience.}
and (iii)
an output scheme to retrieve the data in  an $m$-qubit
superposition state 
$\sum_a |x^{(a)}\ket_D$.
Here, the subscripts $A$ and $D$
stand for the quantum versions of an address register and
a data register, respectively.
Explicitly, a qRAM is
defined as a function
\begin{equation}
\mathrm{qRAM}\colon \sum_{a}|a\ket_A|0\ket_D
\mapsto \sum_a |a\ket_A|x^{{(a)}}\ket_D.
\label{qRAM}
\end{equation}

%%%%%%%%%%%%%%%%%%%%%%%%%%%%%%%%%%%%%%%%%%%%%%%%%%%%%%%%%%%%%%%%%%%%%
\begin{figure}
\centering
\includegraphics[width=0.9\textwidth]{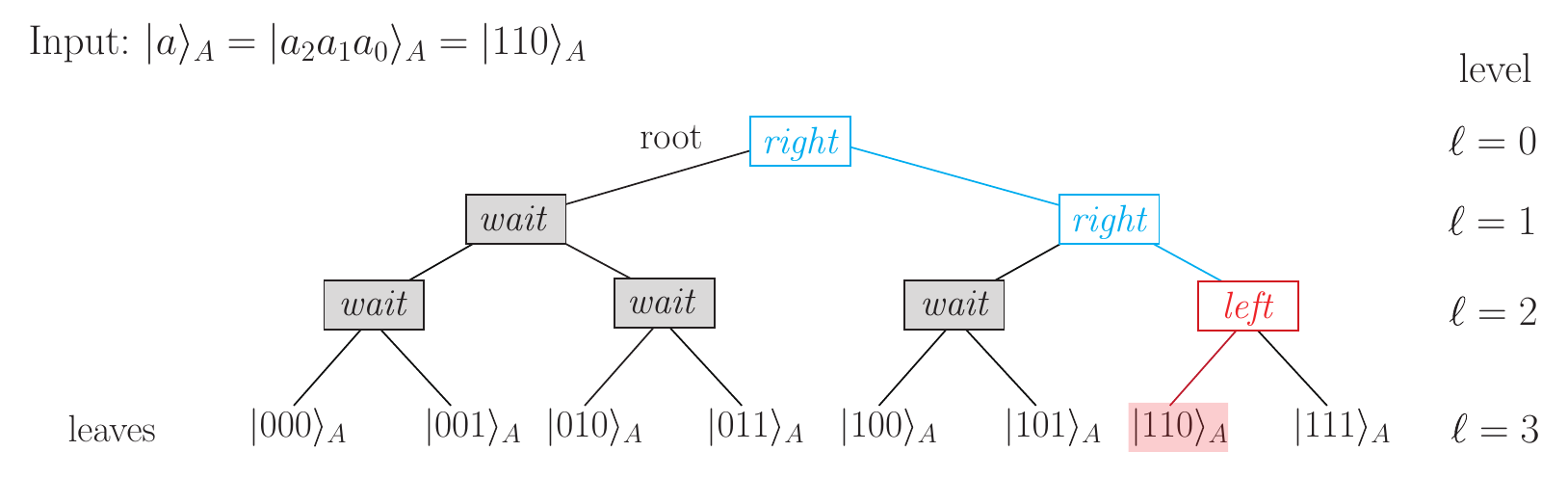}
\caption{The GLM bucket-brigade scheme defined on a binary 
tree with depth $n=3$. A qutrit is installed at each node to route to the
specified memory cells. For instance,  to route to the cell at address $|110\ket_A$, 
the three qutrits must be activated from \textit{wait} to \textit{left}/\textit{right}.
}
\label{bucket}
\end{figure}
%%%%%%%%%%%%%%%%%%%%%%%%%%%%%%%%%%%%%%%%%%%%%%%%%%%%%%%%%%%%%%%%%%%%%
Giovannetti, Lloyd and Maccone (GLM) proposed a remarkable qRAM architecture 
using the so-called bucket-brigade routing scheme
\cite{giovannetti2008quantum,
giovannetti2008architectures}. The GLM architecture
is defined on a perfect binary tree with depth $n$ on 
which  $N=2^n$ data are stored in the memory cells placed on the leaves
of the binary tree  (see Fig.~\ref{bucket}). Each node in the 
tree is equipped with a qutrit with three energy labeled \textit{wait}, 
\textit{left}, and \textit{right}, and all the qutrits are initially in 
the \textit{wait} state. The qutrit acts as a router: the 
value $a_{n-1-\ell}\in\{0,1\}$ ($0\le \ell\le n-1$) 
in the address register 
\eqref{address} is delivered to one of the $2^{\ell}$ nodes at the $\ell$th
level of the binary tree,  and if $a_{n-1-\ell}=0$ (resp. $a_{n-1-\ell}=1$), 
activates the qutrit from \textit{wait} to  
\textit{left} (resp. \textit{right}) to route the subsequent $a_{n-2-\ell}$ 
to one of the two child nodes. After $O(n^2)=O(\log^2 N)$ steps, a unique route is 
assigned from the root to the specified memory cell, as schematically
depicted in Fig.~\ref{bucket}.
A quantum bus then arrives at the cell through the assigned route, the data 
stored in the cell is coherently loaded onto the bus, and the bus loaded with 
the data returns to the root  via the route it came from.
Finally, reverting the activated qutrits to 
\textit{wait}, sequentially from the last level, 
yields output in the r.h.s of \eqref{address}.  For each memory call, 
the overall computational cost and qubit resources required to process $N=2^n$ $m$-qubit data 
are $O(n^2+nm)$ and $O(2^n+m)$, respectively.

It is worth noting that the number of qutrits to be activated is only $O(n)$, 
which drastically reduces a cost of maintaining the quantum coherence 
compared to the fan-out scheme (most commonly used in a classical RAM)
that activates $O(2^n)$ qutrits.
In fact, a high resilience of the bucket-brigade qRAM 
to generic noise has been recently proved in \cite{hann2021resilience}.
The GLM qRAM has been improved and is realized efficiently by quantum 
circuits as in \cite{paler2020parallelizing,arunachalam2015robustness,di2020fault}.
Some experimental implementations have also been proposed in \cite{giovannetti2008architectures,
hong2012robust,kyaw2015scalable,hann2019hardware,PRXQuantum.2.030319}. 

More recently, the authors of the present paper have provided a novel qRAM 
algorithm that works on a perfect binary tree but does not require entanglement 
with any quantum device on the nodes \cite{asaka2021quantum}. In this sense, 
this algorithm promises to reduce the cost of maintaining quantum coherence 
compared to the bucket-brigade scheme, but its implementation has remained 
open until now. The purpose of this paper is 
to physically implement this qRAM algorithm using a multi-particle 
continuous-time quantum walk with two internal states. 

Our qRAM architecture is  roughly sketched as follows. First, quantum information 
is dual-rail encoded into  quantum walkers moving 
on parallel paths; a single-qubit data is represented by the presence of 
a  walker on one of the two parallel paths. Namely, the arbitrary $m$-qubit 
data associated with $n$-qubit address information is represented by a set of 
$n+m$ quantum walkers traveling on half of $2(n+m)$ paths. 
Second, each walker possesses two internal states (e.g., the spin-up and 
down states of an electron). Depending on the internal state, the roundabout gate
allocated at each node of the binary trees passes the walker to one of the two child nodes. 
The address information is sequentially encoded into the internal states 
so that the set of walkers is properly delivered to the designated memory cells.
Finally, the data in the cell is copied  by simply changing
the positions of the walkers in the data register. The set of the walkers
carrying the data is retrieved by the reverse operation of the
routing scheme. 

In the above implementation, the roundabout gate 
can be actually realized by the scattering of the walker from a directed 
graph \cite{asaka2021two}. The encoder, which converts the positional 
information of the path traveled by the specified walkers into the 
internal states of the walkers, is implemented by a combination of 
roundabout gates and single-qubit gates acting on the internal state 
of the walkers.
The main advantages of our architecture are as follows. (i) The processing 
is fully parallelized without using any ancilla qubit,
and can access and retrieve the $m$-qubit data associated
with $n$-qubit address information in $O(n\log(n+m))$ steps. The qubit resources
necessary for the processing is $O(n+m)$.
 (ii) The walkers are not entangled 
with any device on the binary trees,  thus reducing the cost of maintaining the 
quantum coherence. (iii) It does not require any time-dependent control:
the qRAM process is automatically achieved by just passing the walkers through
binary trees. (iv) Using the model developed in the first paper \cite{asaka2021two}, 
it is possible to design a unified universal quantum computer that is 
compatible with the qRAM developed in this paper.

The rest of this paper is outlined as follows. Sec.~2 describes the 
general setup and gives an overview of our qRAM architecture.  Some devices 
developed in the first paper \cite{asaka2021two}, which are required 
in the present paper, are also summarized. A physical implementation of the qRAM 
is provided in Sec.~3. In Sec.~4, an alternative qRAM scheme that transforms 
a trivial state into a superposition of information 
stored in the specified memory cells: 
\begin{equation}
\widetilde{\mathrm{qRAM}}\colon
|0\ket_A |0\ket_D\mapsto \sum_a |a\ket_A|x^{(a)}\ket_D
\end{equation}
(cf. \eqref{qRAM}) is proposed.
The last section is
devoted to the summary and discussion, where we briefly explain how to extract
\textit{quantum} information (i.e. information in quantum superposition) in
the designated cells instead of classical information. 
%%%%%%%%%%%%%%%%%%%%%%%%%%%%%%%%%%%%%%%%%%%%%%%%%%%%%%%%%%%%
\section{Preliminaries}
%%%%%%%%%%%%%%%%%%%%%%%%%%%%%%%%%%%%%%%%%%%%%%%%%%%%%%%%%%%%
This section gives an overview of the qRAM architecture, which
is a physical realization of the algorithm developed in \cite{asaka2021quantum}.
The architecture uses 
some quantum gates implemented by multi-particle continuous-time 
quantum walks \cite{asaka2021two}.
%%%%%%%%%%%%%%%%%%%%%%%%%%%%%%%%%%%%%%%%%%%
\subsection{Setup and layout of the qRAM}
%%%%%%%%%%%%%%%%%%%%%%%%%%%%%%%%%%%%%%%%%%%
%%%%%%%%%%%%%%%%%%%%%%%%%%%%%%%%%%%%%%%%%%%%%%%%%%%%%%%%%%%
\begin{figure}[t]
\centering
\includegraphics[width=0.4\textwidth]{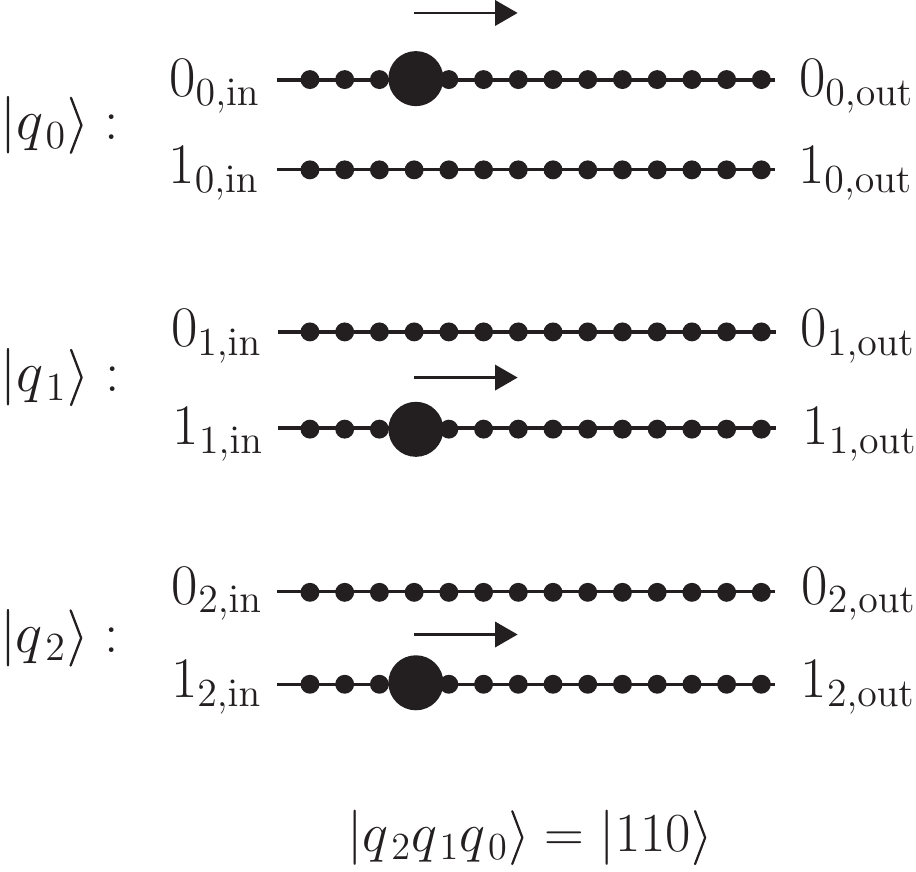}
\caption{A dual-rail encoding of the state $|q_2q_1q_0\ket=|110\ket$. }
\label{dual-rail}
\end{figure}
%%%%%%%%%%%%%%%%%%%%%%%%%%%%%%%%%%%%%%%%%%%%%%%%%%%%%%%%%%%
Our qRAM architecture employs a dual-rail encoding in which  
data and address information are represented as the positions of the
paths the quantum walkers moving; a single-qubit state $|q_j\ket\in\mathbb{C}^2$ is 
expressed by the presence of a quantum walker in one of two parallel 
paths:
\begin{align}
|q_j\ket=\delta_{q_j,0}|2j\ket_\mathrm{p}+\delta_{q_j,1}|2j+1\ket_\mathrm{p} \quad
(0\le j\le n+m-1),
\label{dual1}
\end{align}
where $|2j\ket_\mathrm{p}\in\mathbb{C}^2$ (resp. $|2j+1\ket_\mathrm{p}\in\mathbb{C}^2$) indicates
that a walker is moving on the ($2j$)th (resp. $(2j+1)$th path). 
Correspondingly, an ($n+m$)-qubit state is given by
\begin{equation}
|q_{n+m-1}\cdots q_{0}\ket=
|q_{n+m-1}\ket\otimes \cdots\otimes |q_0\ket
=\bigotimes_{j=0}^{n+m-1}\left(
\delta_{q_j,0}|2j\ket_\mathrm{p}+\delta_{q_j,1}|2j+1\ket_\mathrm{p}
\right)\in(\mathbb{C}^2)^{\otimes(n+m)}.
\label{dual2}
\end{equation}
Fig.~\ref{dual-rail} shows an example of a dual-rail encoded state. 
For our purposes, we assign the first $n$ qubits and the remaining $m$ qubits
to the address and data registers, respectively:
\begin{align}
&|a\ket_A=|a_{n-1}\cdots a_{0}\ket_A=|a_{n-1}\ket_{A_{n-1}}\otimes
\cdots \otimes |a_{0}\ket_{A_0}=|q_{n-1}\cdots q_0\ket\in
(\mathbb{C}^2)^{\otimes n}, \nn \\
&|x^{(a)}\ket_D=|x_{m-1}^{(a)}\cdots x_0^{(a)}\ket_D
=|x_{m-1}^{(a)}\ket_{D_{m-1}}\otimes \cdots \otimes |x_0^{(a)}\ket_{D_0}
=|q_{n+m-1}\cdots q_{n}\ket\in(\mathbb{C}^2)^{\otimes m}.
\end{align}

\begin{figure}
\centering
\includegraphics[width=0.8\textwidth]{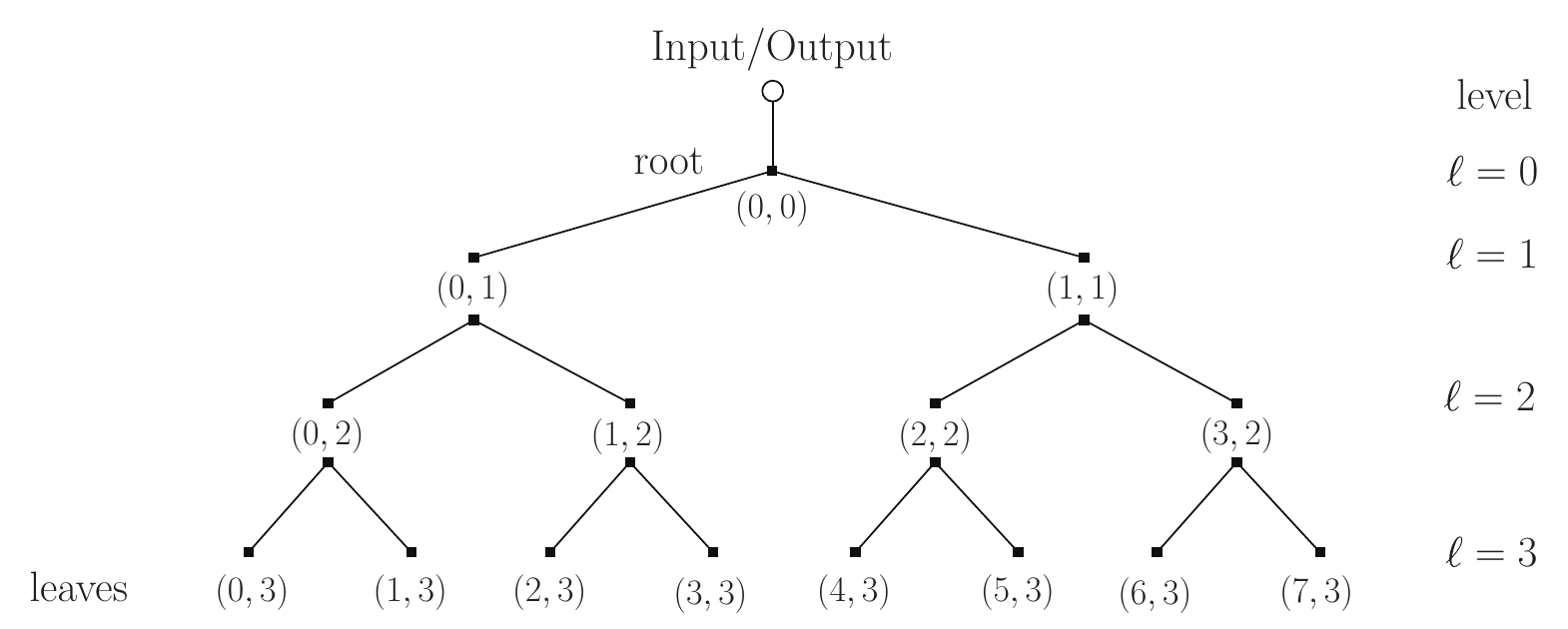}
\caption{A perfect binary tree with depth $n=3$. 
At level $\ell$ ($0\le \ell \le n$) of the binary tree, the $w$th 
node counting from the left is labeled as $(w,\ell)$ ($0\le w \le 2^{\ell}-1$;
$0\le \ell \le n$). Each node $(w,\ell)$ has two child nodes
$(2w,\ell+1)$ (left child) and $(2w+1,\ell+1)$ (right child) for 
$0\le\ell \le n-1$. 
The input/output terminal is connected to the root node $(0,0)$
by a path.
}
\label{binary-tree}
\end{figure}

\begin{figure}
\centering
\includegraphics[width=0.95\textwidth]{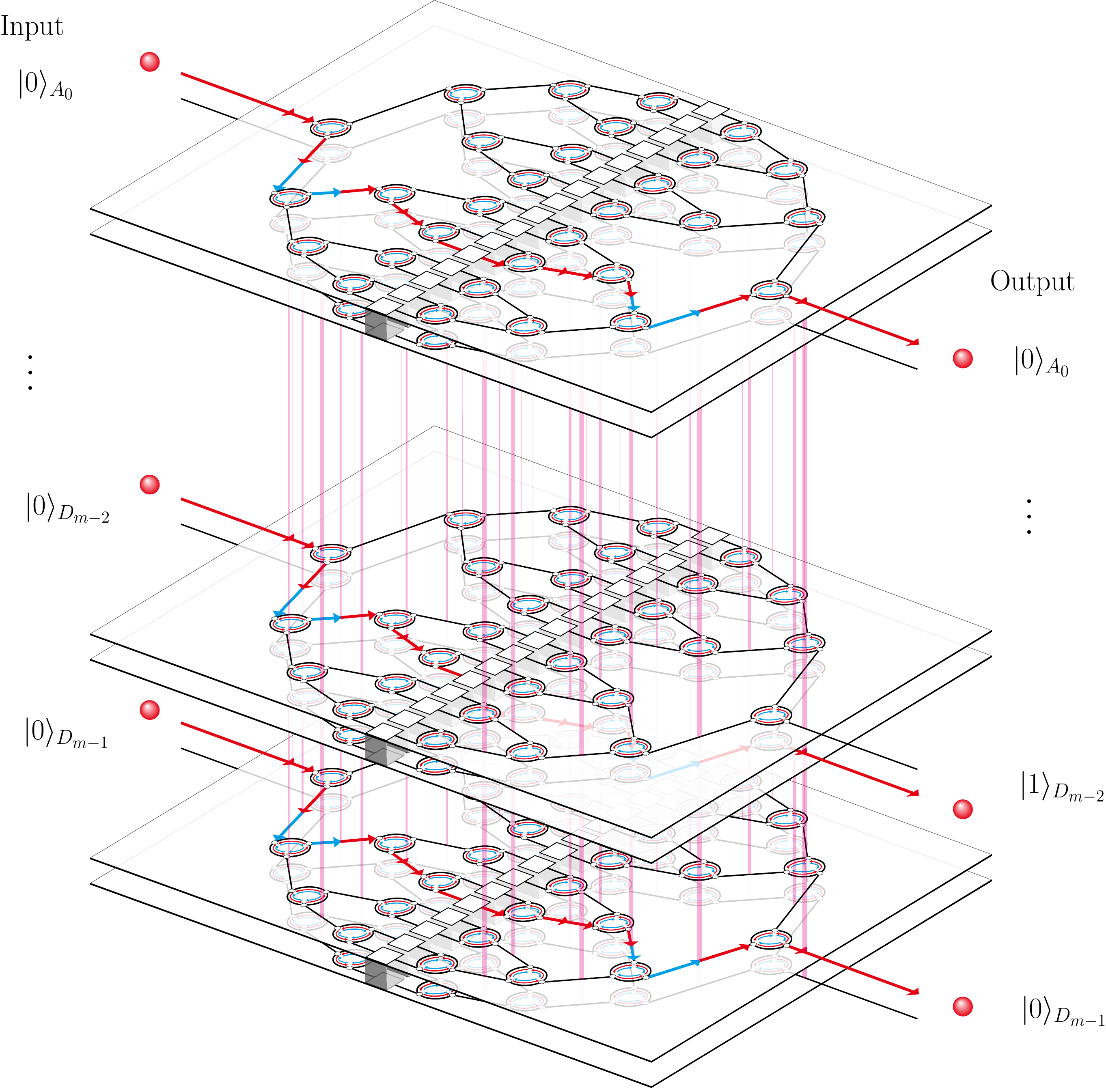}
\caption{An overview of the current qRAM architecture. 
Due to the dual-rail encoding  (see \eqref{dual1}, \eqref{dual2} and 
also Fig.~\ref{dual-rail}),
the architecture is designed on $2(n+m)$ sheets. Each sheet has two 
perfect binary trees of depth $n$ (cf. Fig.~\ref{binary-tree}), 
with their $2\times 2^n$ leaves sandwiching the $2^n$ memory cells.
The $n+m$ red quantum walkers at the left back (resp. right front) 
represent the input (resp. output) state. In the routing scheme, the 
roundabout gate is set up at each node of the trees so that it passes a red 
walker (resp. blue walker) to the left (resp. right) child node. 
The address information $|a_{n-1-\ell}\ket_{A_{n-1-\ell}}$ is encoded in 
the internal state of all the $n+m$ walkers as
$\otimes_{j=0}^{n+m-1}|a_{n-1-\ell}\ket_{C_j}$ while they move to the node $(w,\ell)$
($0\le w \le 2^{\ell}-1$; $0\le \ell \le n-1$) from its parent node 
(the parent node for the root node $(0,0)$ denotes the input terminal). 
This process can be accomplished by 
the device $\mathcal{E}_{(w,\ell)}$ 
intersecting perpendicular to the paths between two levels 
$\ell-1$ and $\ell$ ($0\le \ell\le n-1$) ($\ell=-1$ denotes the input terminals). 
The data $x^{(a)}$ stored in the memory cell at address $a$ is
loaded to the walkers arriving at the cell. By reversing the routing scheme, the walkers
loaded with the data (in superposition) are retrieved as the output.}
\label{overview}
\end{figure}

The $(n+m)$ quantum walkers (in superposition) access the specified memory cell(s) 
through half of the $2(n+m)$ parallel paths and retrieve the data stored in the cell(s). 
To this end, 
we prepare $2(n+m)$ parallel sheets on each of which two perfect binary trees of depth 
$n$ are arranged so that the $2^n$ memory cells are sandwiched between 
the two sets of $2^n$ leaves, as schematically shown in Fig.~\ref{overview}. 
(See also Fig.~\ref{binary-tree} as a detailed description of a perfect binary tree.) 
A set of $n+m$ walkers (possibly in superposition) at input (resp. output) terminals 
of the first (resp. second) binary trees corresponds to the input (resp. output) state. 
Let $(w,\ell)$  ($0\le w \le 2^{\ell}-1$;
$0\le \ell \le n$) be the $w$th node from the left at the $\ell$th level of 
the perfect binary tree, and let $|w,\ell\ket_B\in\mathbb{C}^{2^{n+1}-1}$ 
denote that a set of $n+m$ walkers (called  a ``bus") 
is moving toward the node $(w,\ell)$ from its parent node (the parent 
node for the root node $(0,0)$ denotes the input/output terminal
(see Fig.~\ref{binary-tree})).  Namely, the bus that passes between 
these two nodes, carrying $m$-qubit data $|x^{(a)}\ket_D$ associated 
with an $n$-qubit address $|a\ket_A$, is represented as
 $|a\ket_A|w,\ell\ket_B  |x^{(a)}\ket_D$.
%
%%%%%%%%%%%%%%%%%%%%%%%%%%%%%%%%%%%%%%%%%%%%%%%%%%%%%%%%%
%%%%%%%%%%%%%%%%%%%%%%%%%%%%%%%%%%%%%%%%%%%%%%%%%%%%%%%%
%
%
%%%%%%%%%%%%%%%%%%%%%%%%%%%%%%%%%%%%%%%%%%%%%%%%%%%%%%%%%
%%%%%%%%%%%%%%%%%%%%%%%%%%%%%%%%%%%%%%%%%%%%%%%%%%%%%%%%
%

All the $(n+m)$ quantum walkers possess 
two internal states (e.g., the spin-up and  down states
of an electron). Let 
$|c\ket_{C_j}\in\mathbb{C}^2$ ($c\in\{0,1\}$, $0\le j\le n+m-1$)  be the internal state of 
the $j$th walker, and we call the walker with $|0\ket_{C_j}$ and
$|1\ket_{C_j}$ a ``red walker" and a ``blue walker", respectively. 
In principle, we assume that the internal states are initialized to be 
red $|0\ket_{C_j}$ ($0\le j \le n+m-1$) before  processes.
All the walkers at the input/output terminals are colored red 
($|0\ket_C:=\otimes_{j=0}^{n+m-1}|0\ket_{C_j}\in(\mathbb{C}^2)^{\otimes (n+m)}$), and 
just before or just after passing through the nodes, all they are 
colored red or blue ($|2^{n+m}-1\ket_C:=\otimes_{j=0}^{n+m-1}|1\ket_{C_j}
\in(\mathbb{C}^2)^{\otimes (n+m)}$)
according to the address information \eqref{address}. The address information
 is temporarily encoded in the internal states by a unitary gate $\mathcal{E}_{(w,\ell)}$
that intersects perpendicular to the paths between two levels 
$\ell-1$ and $\ell$ ($0\le \ell\le n-1$) ($\ell=-1$ denotes the input terminals),
as shown in Fig.~\ref{overview}. (See the next section for more details about 
$\mathcal{E}_{(w,\ell)}$.)
A roundabout gate is set up at each node to move a red walker (resp. blue 
walker) to the left child node (resp. right child node) in the routing scheme, 
and do exactly the opposite in the output scheme.
The data stored in the memory cells are loaded to the walkers in the data 
register, which is simply realized by changing their positions, as
described in the subsequent section.
The walkers loaded with the data are
retrieved in the output scheme, which is accomplished by
reversing the routing scheme. That is a layout of our qRAM 
given by a function 

\begin{equation}
\begin{array}{rccc}
\mathrm{qRAM}\colon 
&(\mathbb{C}^2)^{\otimes \{2(n+m)+\log_2(2^{n+1}-1)\}}  
&\longrightarrow
&(\mathbb{C}^2)^{\otimes \{2(n+m)+\log_2(2^{n+1}-1)\}}    \\
& \rotatebox{90}{$\in$}&               & \rotatebox{90}{$\in$} \\
&  \displaystyle{\sum_{a\in\mathscr{A}}|a\ket_A |0,0\ket_B |0\ket_C |0\ket_D}
& \longmapsto   
&\displaystyle{\sum_{a\in\mathscr{A}} |a\ket_A |0,0\ket_B |0\ket_C |x^{{(a)}}\ket_D},
\end{array}
\label{qRAM2}
\end{equation}
where $\mathscr{A}\subset\{0,\cdots,2^n-1\}$ denotes the set of the 
addresses of the specified memory cells.
%%%%%%%%%%%%%%%%%%%%%%%%%%%%%%%%%%%%%%%%%%%
\subsection{Quantum gates}
%%%%%%%%%%%%%%%%%%%%%%%%%%%%%%%%%%%%%%%%%%%
Next, we briefly introduce several elementary quantum gates developed in the 
previous paper \cite{asaka2021two} that are necessary for the design 
of the current qRAM architecture.\\

\noindent
\textbf{(a) Single-qubit gates } Arbitrary single-qubit gates are universally 
realized by a combination of roundabout gates, and rotation gates 
acting on the internal states of the walker. 

The roundabout gate
serves as a router that moves a  walker either clockwise or counterclockwise 
from one path to the next according to the internal state of the walker:
\begin{align}
&U_\mathrm{R}^{(\mathrm{l})}=|0\ket\bra 0|_{C_j} U_{\mathrm{R}}+
  |1\ket\bra 1|_{C_j} U^{\dagger}_{\mathrm{R}},\quad
U_{\mathrm{R}}^{(\mathrm{r})}=U_{\mathrm{R}}^{(\mathrm{l})}{}^{\dagger},\nn \\ 
&U_{\mathrm{R}}=\sum_{k,l=0}^{2}\delta_{l,k+1}|j_l\ket\bra j_k|_\mathrm{p}\quad
(k,l\in\mathbb{Z}/3\mathbb{Z}=\{0,1,2\}).
\label{eq-RA}
\end{align}
Here, $U_\mathrm{R}^{(\mathrm{l})}$ (resp. $U_\mathrm{R}^{(\mathrm{r})}$) is a
unitary operator that moves a 
red walker (a walker with the internal state $|0\ket_{C_j}$) 
(resp. blue walker (a walker with $|1\ket_{C_j}$) ) clockwise (resp. counterclockwise)
to the next path. Graphically, it is represented as
\begin{equation}
\includegraphics[width=0.6\textwidth]{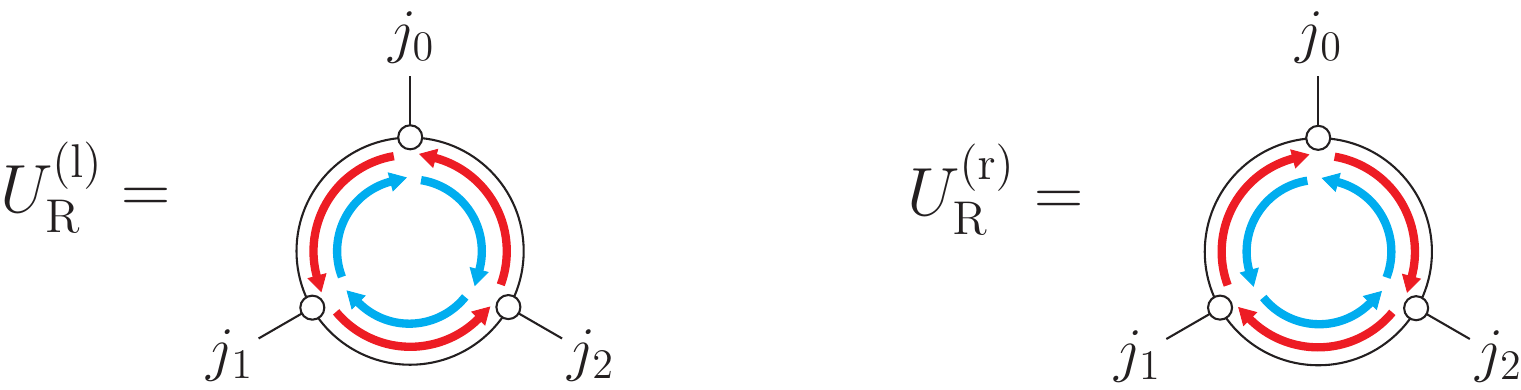} .
\label{fig-RA}
\end{equation}
For example, the motion of a red/blue walker that enters the $U_\mathrm{R}^{(\mathrm{l})}$ or $U_\mathrm{R}^{(\mathrm{r})}$ gate
from path $j_0$ is graphically given
\begin{equation}
\includegraphics[width=0.8\textwidth]{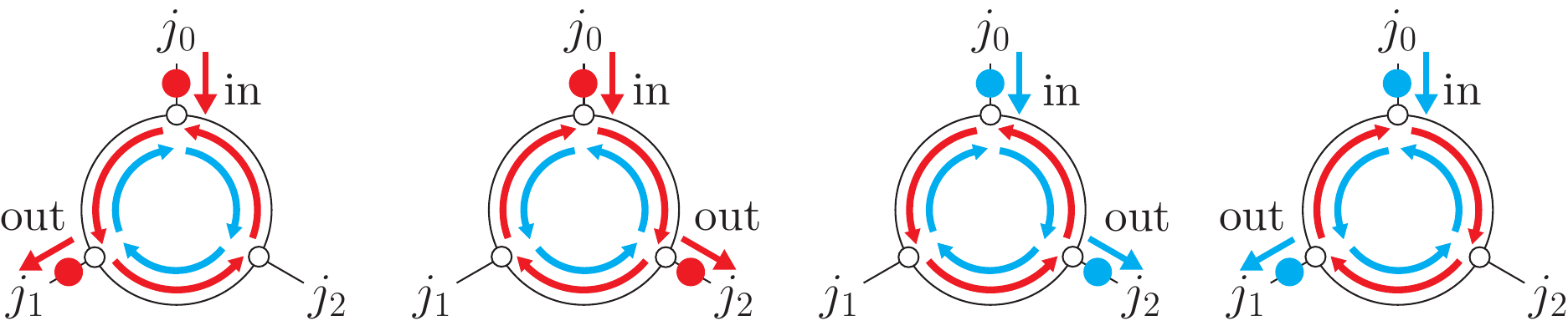} .
\end{equation}
Physically the roundabout gate can be implemented by a single-particle scattering
from a directed graph as shown in Sec. 3 in \cite{asaka2021two}.

Let us pictorially denote a quantum gate $U_{C_j}$ acting on the 
internal state $|c\ket_{C_j}$ of the $j$th quantum walker as
\begin{equation}
\includegraphics[width=0.3\textwidth]{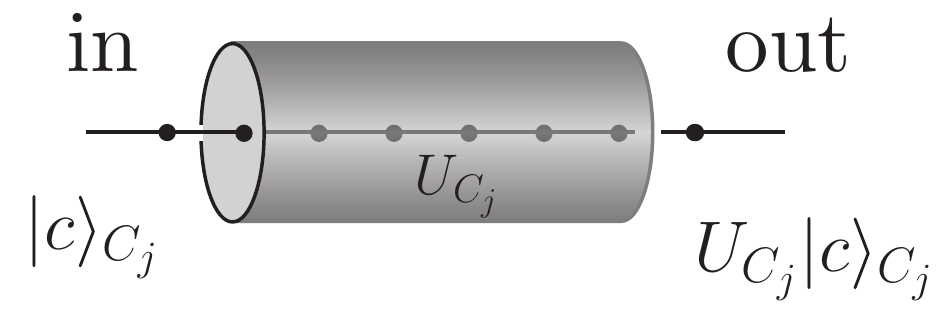} .
\end{equation}
An arbitrary single-qubit gate $U_{C_j}$ is universally realized by
$U_{C_j}=e^{i\theta_0}R_z(\theta_1)R_y(\theta_2)R_z(\theta_3)$
($\theta_k\in\mathbb{R}$ ($k=0,1,2,3$)) \cite{nielsen2002quantum,williams2011quantum}, where
$R_y(\theta):=e^{-i \theta Y /2}$
(resp. $R_z(\theta):=e^{-i \theta Z/2}$) is the operator
that rotates the Bloch vector around the $y$-axis ($z$-axis)
by a given angle $\theta$. For example, the Pauli-X gate is
represented as $X_{C_j}=R_y(\pi)_{C_j}$ whose action on the states 
$|0\ket_{C_j}$ and $|1\ket_{C_j}$ are graphically represented as
\begin{equation}
\includegraphics[width=0.55\textwidth]{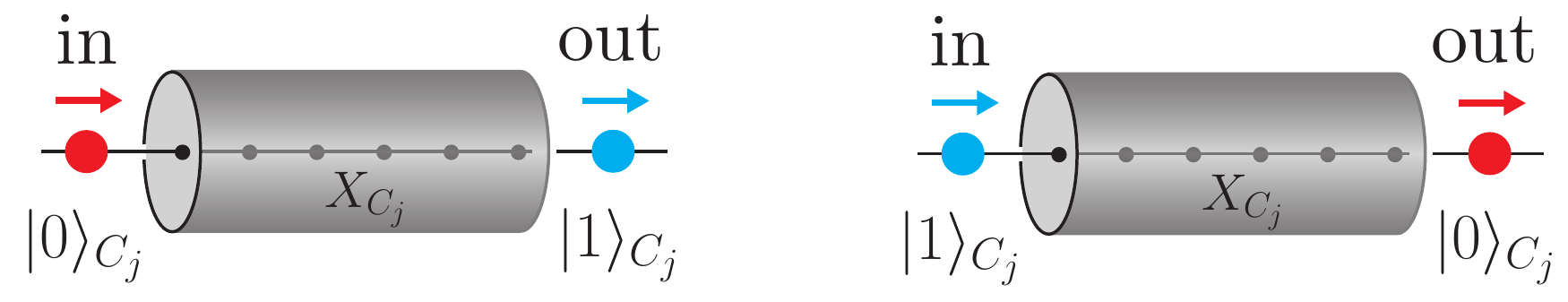} .
\end{equation}
For the spin-1/2 fermionic 
quantum walks, the operator $R_y(\theta)$ 
(resp. $R_z(\theta)$) is physically realized by applying a magnetic field 
$H$ in the direction of $y$-axis (resp. $z$-axis) with a specific strength 
depending on the angle $\theta$. See Fig.~5 in the first paper \cite{asaka2021two}. 

Combining the roundabout gate and the gate $U_{C_j}$, one can construct the single-qubit
gate $U_j$ acting on the state $|q_j\ket$, i.e.,
$
U_j (|0\ket_{C_j}|q_j\ket)=|0\ket_{C_j}(U_j|q_j\ket)
$
as given in \cite{asaka2021two}:
\begin{equation}
\includegraphics[width=0.75\textwidth]{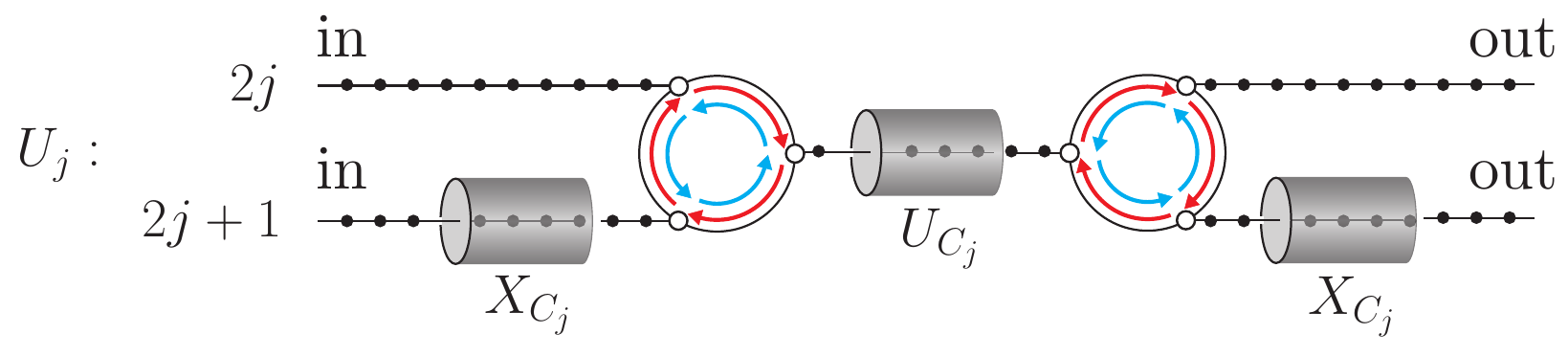},
\label{sqg}
\end{equation}
where a red walker is considered as an input walker, i.e., $|0\ket_{C_j}|q_j\ket$. 
For instance, a walker passing through the Pauli-$X$ gate $X_j$  is depicted
as
\begin{equation}
\includegraphics[width=0.8\textwidth]{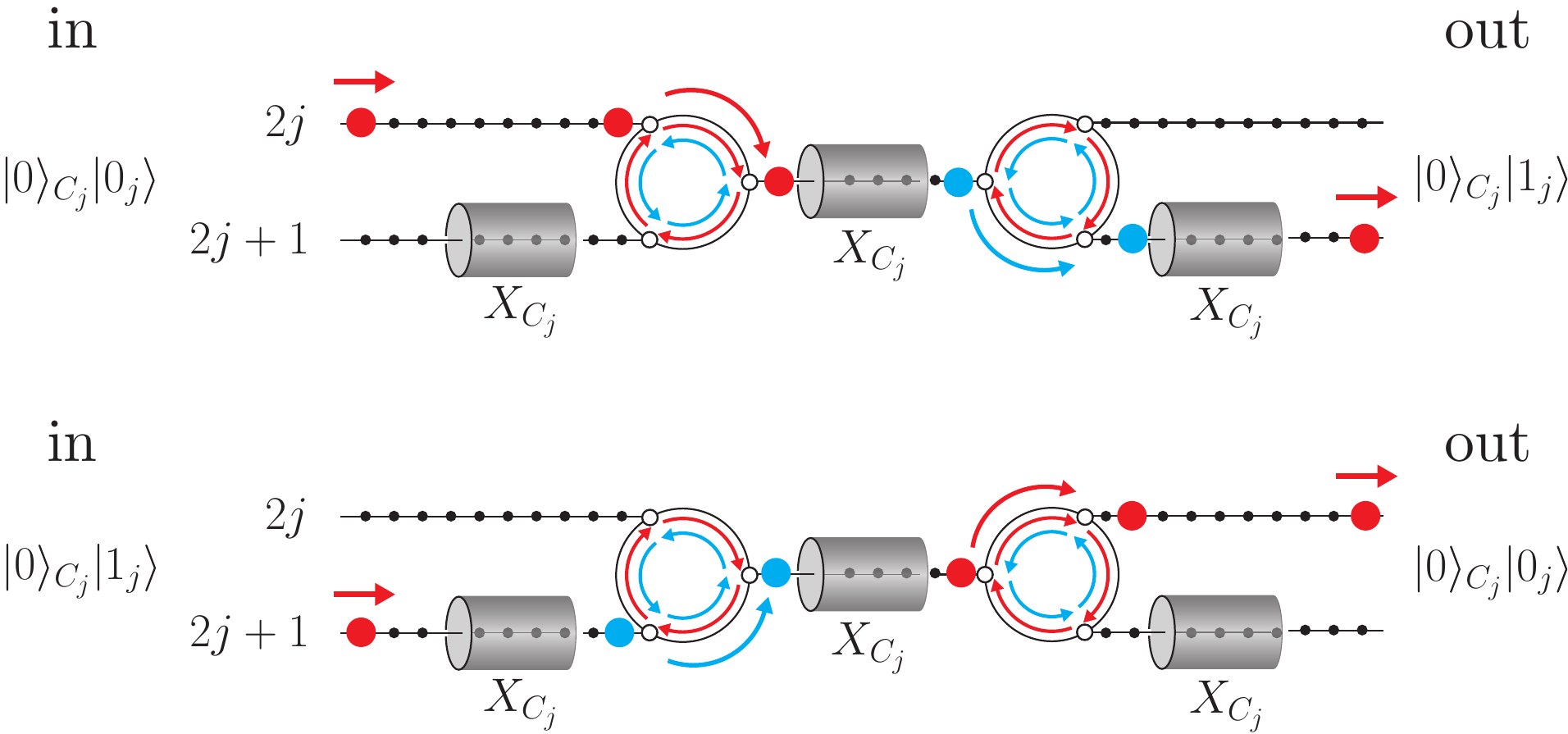}.
\end{equation}
%
%The first part of the above device 
%\begin{equation}
%\includegraphics[width=0.5\textwidth]{UE2.pdf}
%\label{encoder}
%\end{equation}
%acts as an encoder converting the spatial information 
%into the internal state of the quantum walker:
%\begin{equation}
%U_\mathrm{E}: |0\ket_{C_j}|q_j\ket \mapsto
%|q_j\ket_{C_j}.
%\end{equation}

\noindent
\textbf{(b) Two-qubit gates }
Any arbitrary quantum gate can be implemented by a proper
combination of single-qubit gates described above and the
CNOT gate  \cite{nielsen2002quantum,
williams2011quantum}. The CNOT gate $\mathrm{CX}_{jk}$
acting  non-trivially on
$|q_{j}\ket\otimes|q_{k}\ket$ is decomposed to
\begin{equation}
\mathrm{CX}_{j k}=
H_{k}\mathrm{CP}_{j k}H_{k},
\label{CNOT}
\end{equation}
where $H_{k}(=i R_y(\pi/2)_k R_z(\pi)_k)$ is the Hadamard gate acting on $|q_{k}\ket$,
which is achieved by setting $U_{C_k}=H_{C_k}$ in 
\eqref{sqg}. 
$\mathrm{CP}_{j k}$ is a controlled phase
gate 
\begin{equation}
\mathrm{CP}_{jk}=
\begin{pmatrix}
1 & 0& 0& 0 \\
0 & 1& 0& 0 \\
0 & 0& 1& 0 \\
0 & 0& 0& -1
\end{pmatrix}_{jk},
\label{CP-eq}
\end{equation}
which is physically realized by the scattering of two walkers with the same internal 
state on an infinite path \cite{asaka2021two}. 
(See Sec.~4 in \cite{asaka2021two} for another controlled phase gate):
\begin{equation}
\includegraphics[width=0.4\textwidth]{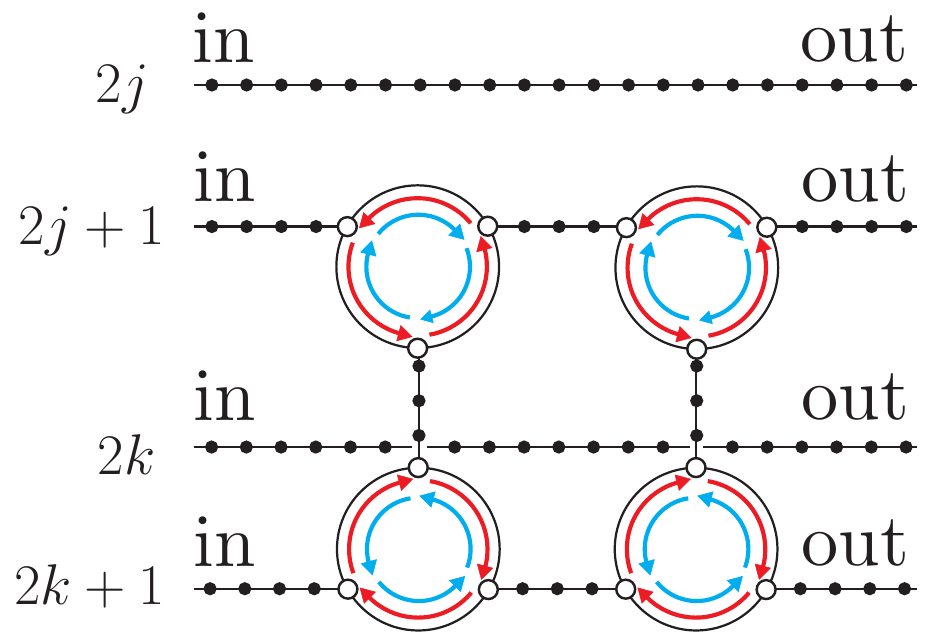},
\end{equation}
where the input state is assumed to be $(|0\ket_{C_k}|q_k\ket)\otimes(|0\ket_{C_j}|q_j\ket)$.
%
%%%%%%%%%%%%%%%%%%%%%%%%%%%%%%%%%%%%%%%%%%%%%%%%%%%%%%%%
\section{Physical implementation of qRAM}
%%%%%%%%%%%%%%%%%%%%%%%%%%%%%%%%%%%%%%%%%%%%%%%%%%%%%%%%
Now we describe an implementation of the qRAM that realizes an algorithm
formulated in \cite{asaka2021quantum}. Let us explain the 
details in the order of (i) the routing scheme $\mathcal{F}$, (ii) the querying 
scheme $\mathcal{Q}$ and (iii) the output scheme $\mathcal{F}^{\dagger}$.
Our qRAM architecture is implemented by these schemes:
\begin{equation}
\text{qRAM}=\mathcal{F}^{\dagger}\mathcal{Q}\mathcal{F}.
\end{equation}

\noindent
\textbf{(i) Routing scheme $\mathcal{F}$ } The routing scheme is a scheme to deliver the 
$(n+m)$ quantum walkers (in superposition) to the desired memory cell(s):
\begin{equation}
\mathcal{F}\colon 
\sum_{a\in\mathscr{A}}|a\ket_A |0,0\ket_B |0\ket_C|0\ket_D\mapsto
\sum_{a\in\mathscr{A}}|a\ket_A |a,n\ket_B |0\ket_{C}|0\ket_D.
\label{routing}
\end{equation}
The input state 
\begin{equation}
\sum_{a\in\mathscr{A}}|a\ket_A|0,0\ket_B|0\ket_C |0\ket_D=
\sum_{a\in\mathscr{A}}|a_{n-1}\cdots a_0\ket_A|0,0\ket_B|0\ket_C |0\ket_D
\end{equation}
is dual-rail encoded into the  positions of the $(n+m)$ red quantum
walkers at input terminals as in Fig.~\ref{dual-rail} and Fig.~\ref{overview}. 
The $(n+m)$ walkers start moving simultaneously toward  leaves. 

The roundabout gate $U_\mathrm{R}^{(\mathrm{l})}$
is installed at each node $(w,\ell)$  ($0\le w \le 2^{\ell}-1$; $0\le \ell \le n-1$)
so that it routes the red walkers (resp. blue walkers) to the 
left (resp. right) child node $(2w,\ell+1)$ (resp. $(2w+1,\ell+1)$):
\begin{equation}
\includegraphics[width=0.85\textwidth]{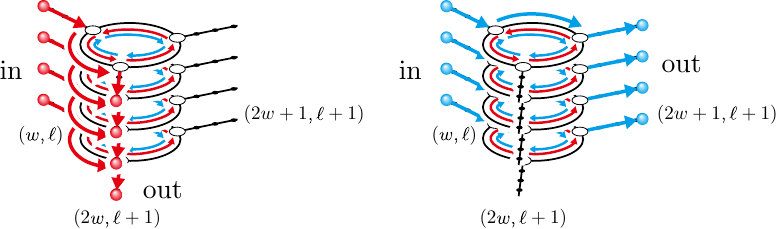}.
\end{equation}
Formally this process is given by the operator $\mathcal{R}_{(w,\ell)}$:
\begin{equation}
\mathcal{R}_{(w,\ell)}\colon
|w,\ell \ket_B\otimes 
\bigotimes_{j=0}^{n+m-1} |c\ket_{C_j}
\mapsto |2w+c,\ell+1\ket_B\otimes
\bigotimes_{j=0}^{n+m-1} |c\ket_{C_j}\quad (c\in\{0,1\}).
\label{shift}
\end{equation}

\begin{figure}
\centering
\includegraphics[width=0.98\textwidth]{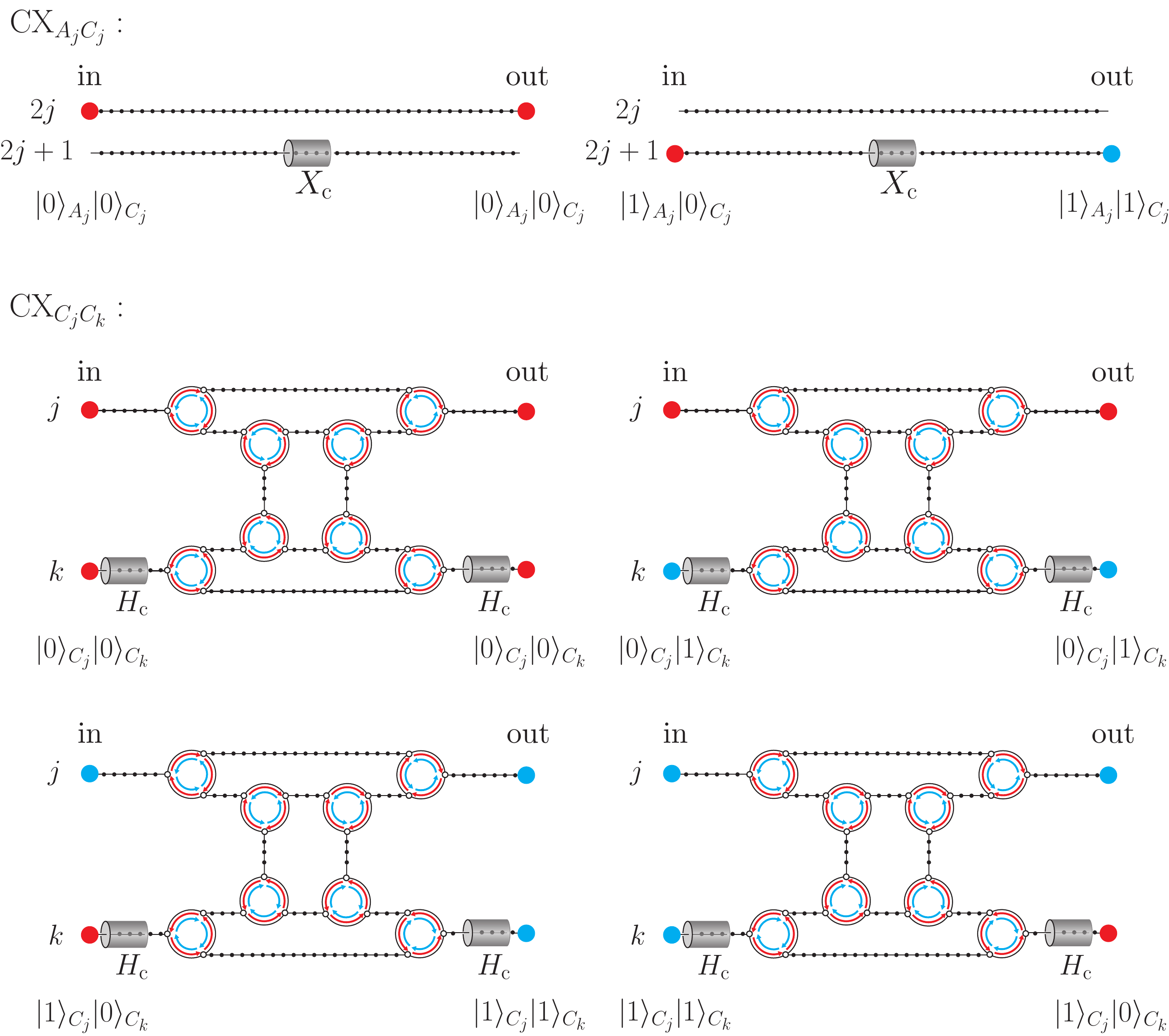}
\caption{
A schematic description of the output states through the 
gates $\mathrm{CX}_{A_jC_j}$  and 
$\mathrm{CX}_{C_jC_k}$ and the corresponding input states.
}
\label{cnotc-pic}
\end{figure}

\begin{figure}
\centering
\includegraphics[width=0.7\textwidth]{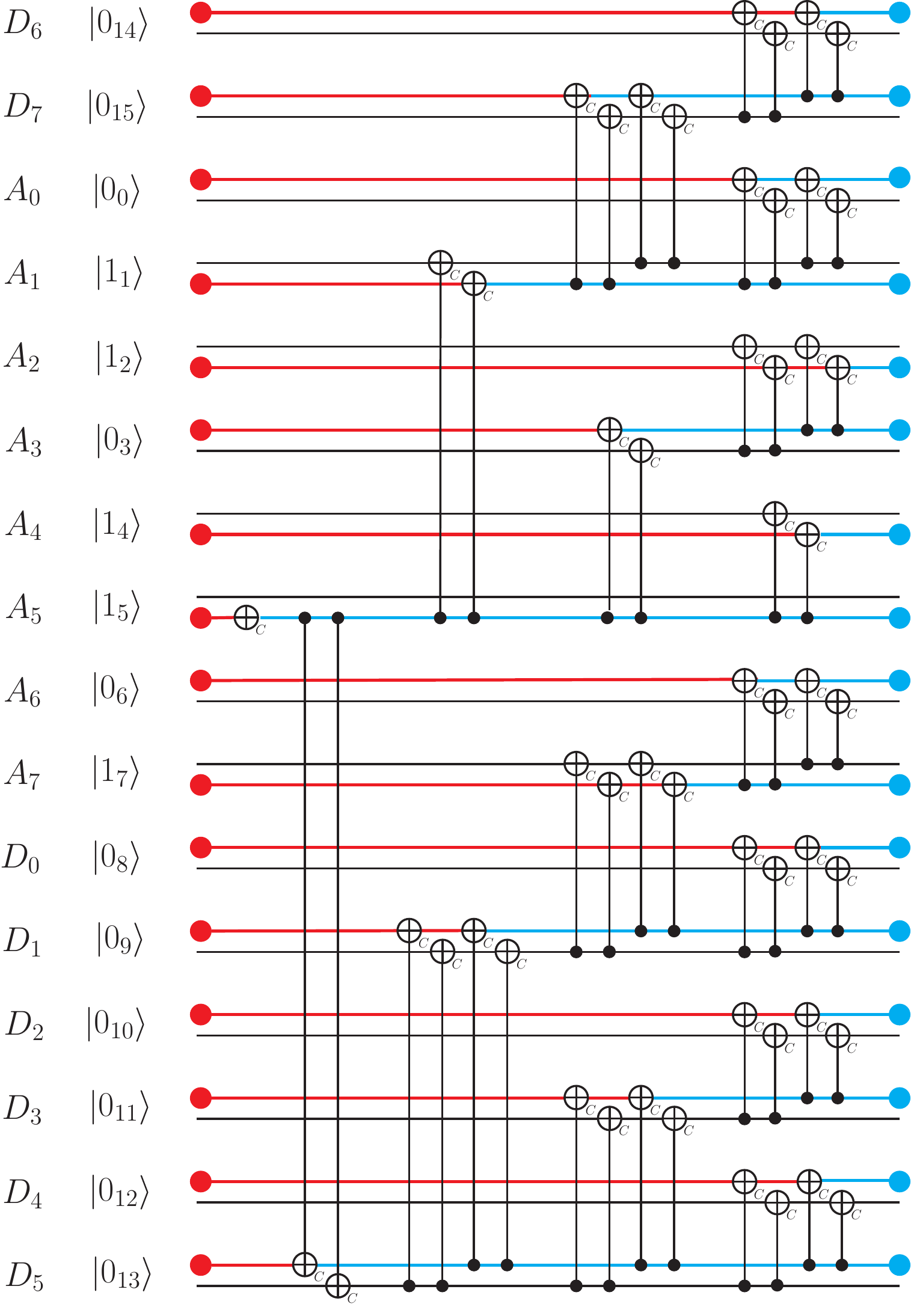}
\caption{A pictorial representation of the action of 
$\mathcal{E}_{(w,\ell)}$
defined by \eqref{encoder} (see also \eqref{encoder-def})
for $w\in2\mathbb{Z}$, $\ell=2$, $n=8$, $m=8$, $p=\log_2(n+m)=4$, 
$|10110110\ket_A$ and $|00000000\ket_D$. After 
$O(p)$ steps, the positional information 
$|a_{n-1-\ell}\ket_{A_{n-1-\ell}}=|1\ket_{A_5}$ is
encoded to the colors
of all the walkers: $\bigotimes_{j=0}^{15}|0\ket_{C_j}
\mapsto \bigotimes_{j=0}^{15}|1\ket_{C_j}$. The number of devices
necessary for the processing is $O(n+m)$. Note that the colors of the
walkers are mixed 
only during this encoding process, otherwise all they are set to either 
red or blue.}
\label{CW-pic}
\end{figure}

The internal states of all the walkers moving to  the node $(w,\ell)$ 
($0\le w \le 2^{\ell}-1$; $0\le \ell \le n-1$) must be $|0\ket_{C_j}$  
(resp. $|1\ket_{C_j}$) ($0\le j\le n+m-1$) for $a_{n-1-\ell}=0$ 
 (resp. $a_{n-1-\ell}=1$), so that the walkers passing through 
the routers at $(w,\ell)$ move to the 
left (resp. right) node.  Namely, the positional information of the path 
traveled by the ($n-1-\ell$)th walker should be encoded to the internal states of all the walkers. 
This encoding process is formally written by the operator $\mathcal{E}_{(w,\ell)}$
($0\le \ell \le n-1$):
\begin{equation}
\mathcal{E}_{(w,\ell)}\colon \bigotimes_{j=0}^{n-1}|a_j\ket_{A_j} \otimes \bigotimes_{j=0}^{n+m-1} 
|w \bmod 2\ket_{C_j}\mapsto
\bigotimes_{j=0}^{n-1}|a_j\ket_{A_j} \otimes \bigotimes_{j=0}^{n+m-1} 
|a_{n-1-\ell}\ket_{C_j}.
\label{encoder-def}
\end{equation}
As shown immediately below, the operator $\mathcal{E}_{(w,\ell)}$ is achieved by
a CNOT gate $\mathrm{CX}_{A_j C_j}$
$(0\le j \le n-1)$:
\begin{equation}
\mathrm{CX}_{A_j C_j}\colon  |a_{j}\ket_{A_{j}}\otimes |c_j\ket_{C_{j}}
\mapsto |a_{j}\ket_{A_{j}}\otimes \left(\delta_{a_j,0}|c_j\ket_{C_j}+\delta_{a_j,1}
X_{C_j}|c_j\ket_{C_j}\right),
\label{enc}
\end{equation}
and a multiple actions of a CNOT gate 
$\mathrm{CX}_{C_j C_k}$
($k\ne j$, $0\le j\le n+m-1$) defined as
\begin{equation}
\mathrm{CX}_{C_j C_k}\colon
|c_j\ket_{C_j}\otimes |c_k\ket_{C_k} \mapsto
|c_j\ket_{C_j}\otimes \left(\delta_{c_j,0}|c_k\ket_{C_k}+\delta_{c_j,1}
X_{C_k}|c_k\ket_{C_k}\right).
\label{cnotc}
\end{equation}
Their graphical representations are, respectively, given by
\begin{equation}
\includegraphics[width=0.6\textwidth]{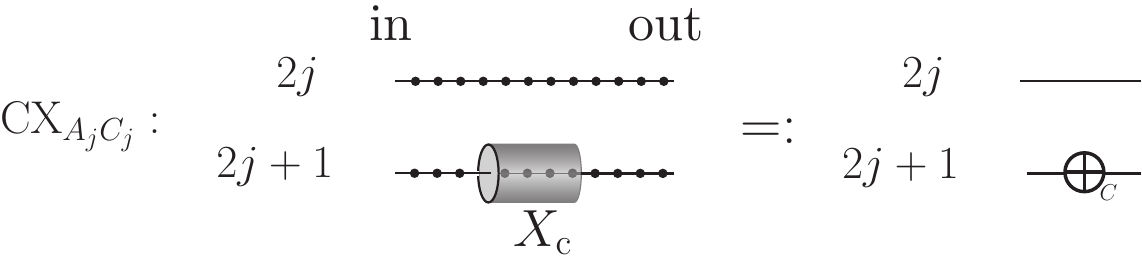},
\end{equation}
and 
\begin{equation}
\includegraphics[width=0.94\textwidth]{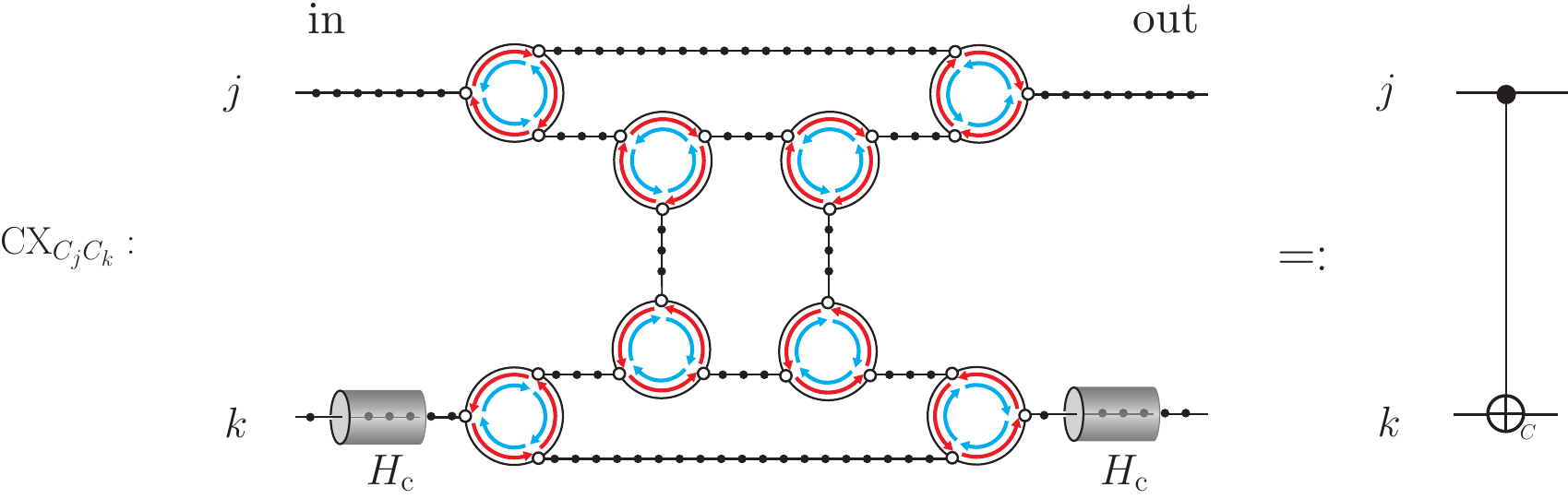}.
\label{cnotc2}
\end{equation}
Also see Fig.~\ref{cnotc-pic}.
%
%%%%%%%%%%%%%%%%%%%%%%%%%%%%%%%%%%%%%%%%%%%%%%%%%%%%%%%%%
%%%%%%%%%%%%%%%%%%%%%%%%%%%%%%%%%%%%%%%%%%%%%%%%%%%%%%%%
%
To implement $\mathcal{E}_{(w,\ell)}$ explicitly,  here and in what follows, 
we assume that the periodic boundary conditions, i.e., 
\begin{equation}
q_{j+n+m}=q_j, \quad C_{j+n+m}=C_j
\end{equation} are imposed,
or equivalently, the particle number $j$ is interpreted as 
\begin{equation}
j\in\{0,\cdots,n+m-1\}
= \mathbb{Z}/(n+m)\mathbb{Z}.
\end{equation}
In addition, we also assume that 
\begin{equation}
n+m=2^{p}, \quad p\in\mathbb{Z}_{\ge 0}
\end{equation}
for simplicity. (An extension to generic $n$ and $m$ is straightforward.)
Under these assumptions, the operator $\mathcal{E}_{(w,\ell)}$
is decomposed to
\begin{align}
&\mathcal{E}_{(w,\ell)}=\mathcal{E}_{\ell}\mathcal{X}_{(w,\ell)},
\qquad \mathcal{X}_{(w,\ell)}:=\prod_{j=0}^{n+m-1}
(\delta_{w\in 2\mathbb{Z}}+\delta_{w\in2\mathbb{Z}+1} X_{C_j}), \nn \\
&\mathcal{E}_{\ell}:=
\mathcal{E}^{(p|p-1)}_{\ell}\cdots\mathcal{E}^{(2|1)}_{\ell}
\mathcal{E}^{(1|0)}_{\ell}\mathrm{CX}_{A_{n-1-\ell}C_{n-1-\ell}},
\label{encoder}
\end{align}
where the operator $\mathcal{X}_{(w,\ell)}$ resets the color 
of the blue walkers to red, $\mathrm{CX}_{A_{n-1-\ell}C_{n-1-\ell}}$ 
is the encoder \eqref{enc} that encodes the information 
$a_{n-1-\ell}$ into the internal state of the 
$(n-1-\ell)$th walker as $|a_{n-1-\ell}\ket_{C_{n-1-\ell}}$, and 
$\mathcal{E}^{(k+1|k)}_{\ell}$ is constructed by
\begin{align}
&\mathcal{E}^{(1|0)}_{\ell}=\left[r,r+2^{p-1}\right], \nn \\
&\mathcal{E}^{(k+1|k)}_{\ell}=
\prod_{s=0}^{2^{k-1}-1}\left[r-s 2^{p-k},r-s 2^{p-k}-2^{p-k-1}\right]\nn \\
&\qquad \qquad \qquad \qquad \quad
\times \left[r+2^{p-1}-s 2^{p-k},r+2^{p-1}-s 2^{p-k}-2^{p-k-1}\right]
\quad (k\ge 1).
\end{align}
Here, $r:=n-1-\ell$ and we have used the abbreviation
\begin{equation}
[j,k]:=\mathrm{CX}_{C_jC_k}, \quad j,k\in\mathbb{Z}/(n+m)\mathbb{Z}
\end{equation}
to simplify the notation.
This process may be intuitively understood by the  graphical representation
as in Fig.~\ref{CW-pic}. Note here that the colors of the
walkers are mixed only during this encoding process, otherwise all they are 
set to either red or blue.
%
%%%%%%%%%%%%%%%%%%%%%%%%%%%%%%%%%%%%%%%%%%%%%%%%%%%%%%%%%
%%%%%%%%%%%%%%%%%%%%%%%%%%%%%%%%%%%%%%%%%%%%%%%%%%%%%%%%
%
Thus, the quantum walkers  appropriately move to the paths
connecting two nodes $(w,\ell)$ and $(2w+a_{n-1-\ell},\ell+1)$ 
($0\le w \le 2^{\ell}-1$; $0\le \ell \le n-1$) by
\begin{align}
&\mathcal{F}^{(\ell+1|\ell)}:=\sum_{w=0}^{2^\ell-1}
\mathcal{R}_{(w,\ell)}\mathcal{E}_{(w,\ell)},\nn \\
&\mathcal{F}^{(\ell+1|\ell)}\colon
|a\ket_A |w,\ell\ket_B\otimes\bigotimes_{j=0}^{n+m-1}|a_{n-\ell}\ket_{C_j}\otimes|0\ket_D \nn \\
&\qquad  \qquad \qquad \mapsto |a\ket_A |2w+a_{n-1-\ell},\ell+1\ket_B 
\otimes \bigotimes_{j=0}^{n+m-1}|a_{n-1-\ell}\ket_{C_j}\otimes|0\ket_D,
\end{align}
where $a_n:=0$.
Recursivelly applying $\mathcal{F}^{(\ell+1|\ell)}$
to the walkers that started moving toward $(w,\ell)$ from
its parent node and finally resetting the colors of the walkers to red,
namely, performing the operator
\begin{equation}
\mathcal{F}=\sum_{w=0}^{2^n-1}\mathcal{X}_{(w,n)}
\mathcal{F}^{(n|n-1)}\cdots \mathcal{F}^{(2|1)}\mathcal{F}^{(1|0)},
\end{equation}
we properly deliver the walkers (in superposition) to 
the designated memory cells, as given by \eqref{routing}.
The depth of the circuit required for the routing scheme is $O(n p)=O(n \log(n+m))$.
\\

\noindent
\textbf{(ii) Querying scheme $\mathcal{Q}$ } 
The querying scheme $\mathcal{Q}$ is a scheme that loads the data $x^{(a)}$ stored 
in the memory cell at the address $a$:
\begin{equation}
\mathcal{Q}\colon
\sum_{a\in\mathscr{A}}|a\ket_A |a,n\ket_B |0\ket_C|0\ket_D
\mapsto
\sum_{a\in\mathscr{A}}|a\ket_A |a,n\ket_B |0\ket_C|x^{(a)}\ket_D,
\label{query}
\end{equation}
which is formally realized by 
\begin{equation}
\mathcal{Q}=\sum_{a\in\mathscr{A}}|a,n\ket\bra a,n|_B\otimes
\bigotimes_{i=0}^{m-1} \left(X_{D_i}\right)^{x_i^{(a)}}.
\label{querying}
\end{equation}
%
%%%%%%%%%%%%%%%%%%%%%%%%%%%%%%%%%%%%%%%%%%%%%%%%%%%%%%%%%
\begin{figure}
\centering
\includegraphics[width=0.95\textwidth]{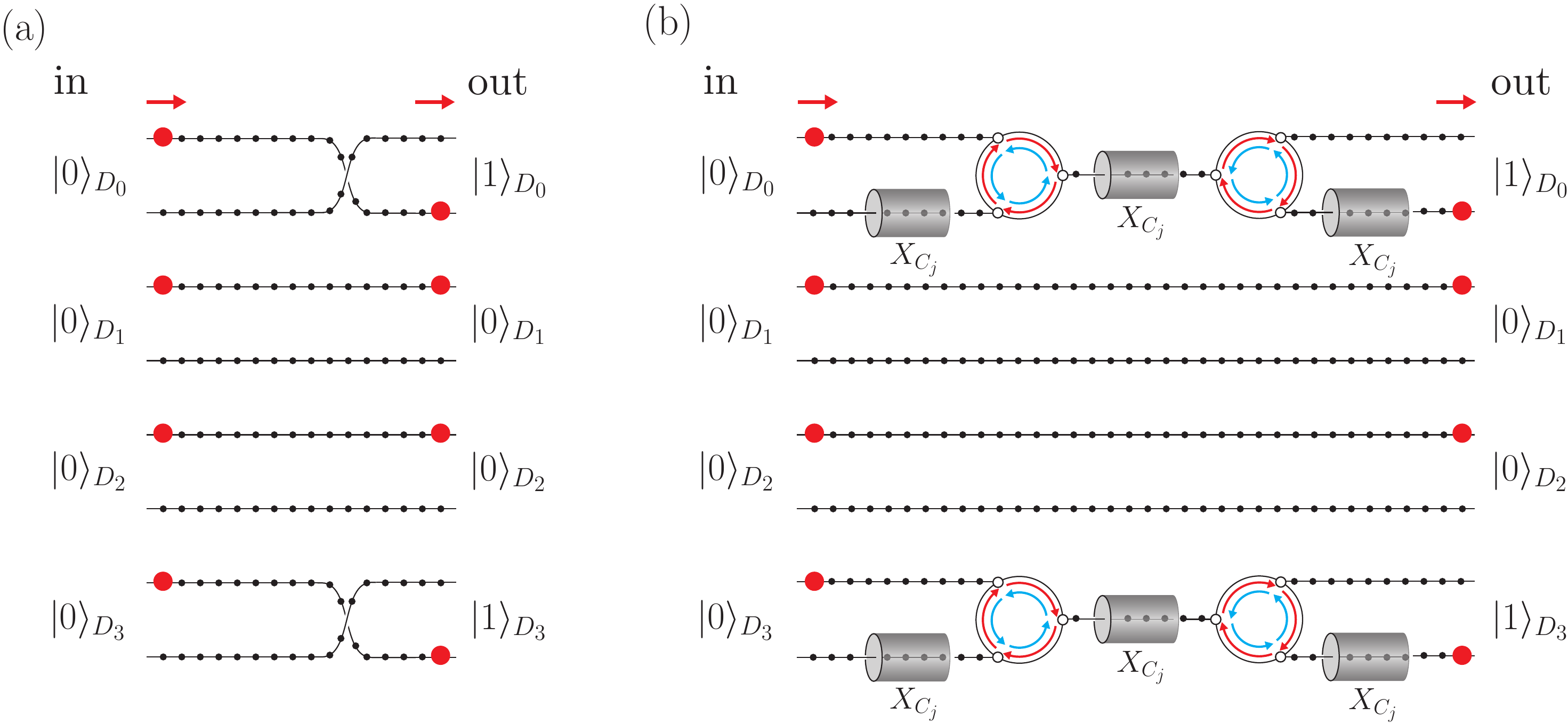}
\caption{A pictorial representation of the querying scheme \eqref{query}
for $|0000\ket_D\mapsto|1001\ket_D$, which can be achieved by (a) exchanging paths or (b) placing the Pauli-X gates.}
\label{query-pic}
\end{figure}
%%%%%%%%%%%%%%%%%%%%%%%%%%%%%%%%%%%%%%%%%%%%%%%%%%%%%%%%
In our architecture, this scheme is implemented by simply exchanging the 
appropriate paths in the data register
or alternatively by placing the Pauli-X gates (see \eqref{sqg}), as pictorially
shown in Fig.~\ref{query-pic}. \\

\noindent
\textbf{(iii) Output scheme $\mathcal{F}^{\dagger}$ } The output scheme is a procedure to
retrieve the data in superposition. In the current approach, this
scheme is achieved by just applying the reverse operation of the
routing scheme, i.e.,
\begin{equation}
\mathcal{F}^{\dagger}\colon
\sum_{a\in\mathscr{A}}|a\ket_A |a,n\ket_B
|0\ket_C|x^{(a)}\ket_D \mapsto
\sum_{a\in\mathscr{A}}|a\ket_A |0,0\ket_B |0\ket_C|x^{(a)}\ket_D.
\label{output}
\end{equation}
As shown in Fig.~\ref{overview}, the scheme can be implemented simply by 
arranging all devices used in the routing scheme so that their positions 
are perfect mirror images across the memory cells (without changing the 
direction of the arrows on the roundabout gates).

In summary, the present architecture processes  $2^n$ $m$-qubit data
in $O(n \log(n+m))$ steps, which requires $O(n+m)$ qubit resources and
$O((n+m) 2^n)$ quantum devices. Table~\ref{resources} compares the number 
of computational steps and quantum resources required for the 
present qRAM architecture and the original bucket-brigade architecture.
Compared to the original bucket-brigade qRAM, the advantages of our architecture 
are that it requires fewer computation steps and qubit resources and does not 
require time-dependent control. On the other hand, the trade-off for these
advantages is that it requires more space and quantum gates, as shown in 
Figs.~\ref{overview} and \ref{CW-pic}.
\begin{table}[htb]
\centering
  \begin{tabular}{|c|c|c|c|} \hline
 method   & \#computational steps & \#qubits & \#quantum gates  \\ \hline \hline
 quantum walk   & $O(n\log(n+m))$  & $O(n+m)$ & $O((n+m)2^n)$  \\ \hline 
 bucket-brigade \cite{giovannetti2008quantum,giovannetti2008architectures}
                & $O(n^2+n m))$ & $O(2^n+m)$ & $O(2^n)$ \\ \hline
  \end{tabular}
\caption{Comparison of the number of computational steps and quantum resources
required for the present method and the original bucket-brigade method.}
\label{resources}
\end{table}

%%%%%%%%%%%%%%%%%%%%%%%%%%%%%%%%%%%%%%%%%%%%%%%%%%%%%%%%
\section{Some modification of routing scheme}
%%%%%%%%%%%%%%%%%%%%%%%%%%%%%%%%%%%%%%%%%%%%%%%%%%%%%%%%
%
By definition \eqref{qRAM} of qRAM, the address information in superposition, 
namely,  $\sum_a |a\ket_A$, is prepared beforehand, with no mention of how 
it is actually constructed. Here, we propose an alternative qRAM architecture 
that transforms a trivial state into a superposition of information 
stored in the desired memory cells:
\begin{equation}
\widetilde{\mathrm{qRAM}}\colon |0\ket_A |0,0\ket_B |0\ket_C |0\ket_D
\mapsto \frac{1}{\sqrt{\mathscr{|A|}}}
\sum_{a\in\mathscr{A}} |a\ket_A |0,0\ket_B |0\ket_C |x^{{(a)}}\ket_D
\label{qRAM3}
\end{equation}
(cf. eq.~\eqref{qRAM2}), which is accomplished by modifying the routing and 
querying scheme slightly. Note that, in \eqref{qRAM3},  the normalization factor
$1/\sqrt{\mathscr{|A|}}$ is written down explicitly to improve 
the perspective of the discussion here.

\begin{figure}
\centering
\includegraphics[width=0.95\textwidth]{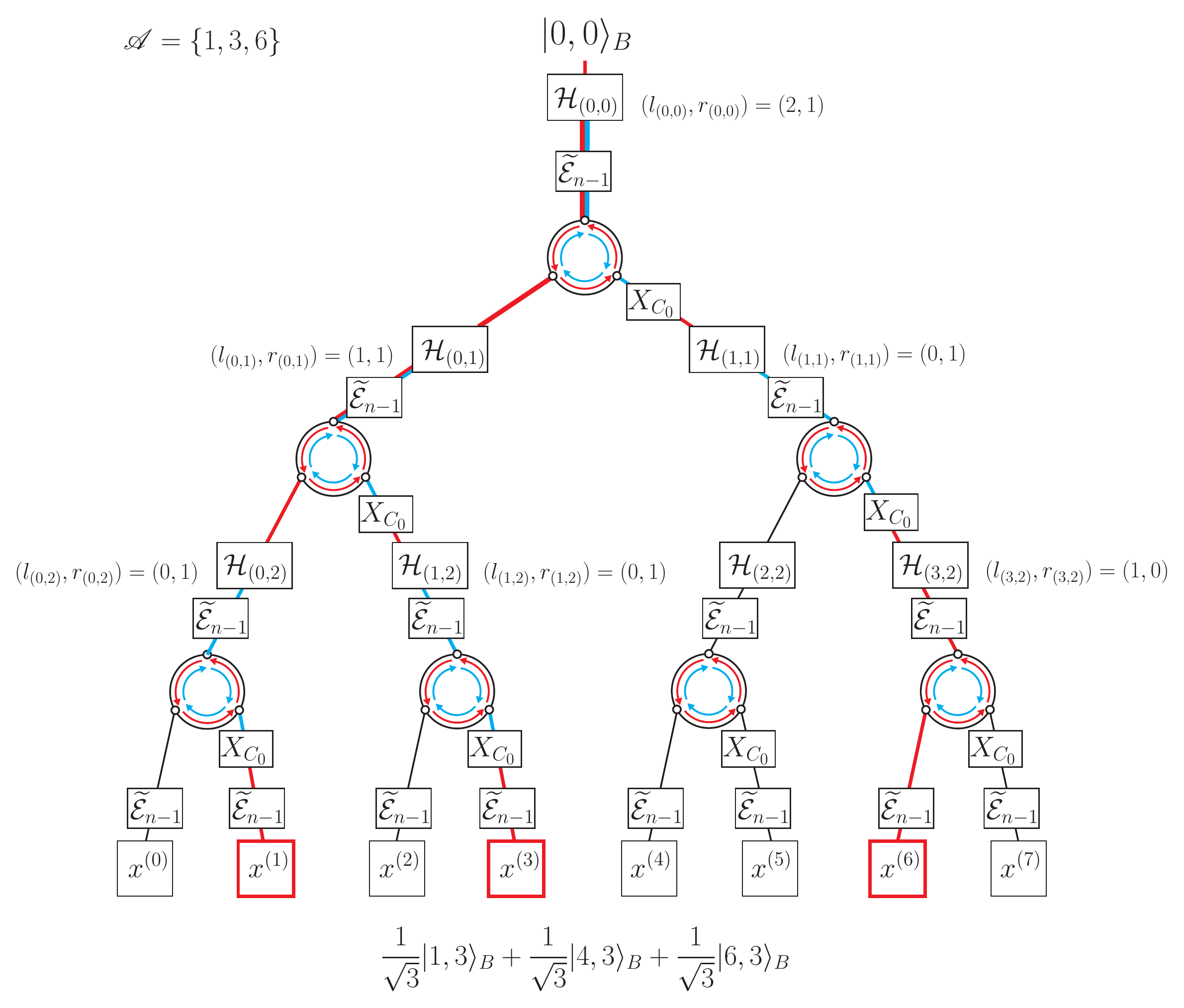}
\caption{A graphical representation of the modified routing scheme $\widetilde{\mathcal{F}}$
defined in \eqref{modified-routing} on the top sheet for $\mathscr{A}=\{1,3,6\}=\{001,011,110\}$.
On the top sheet, the Hadamard-like gate $\mathcal{H}_{(w,\ell)}$ (eq.~\eqref{Hadamard})
is equipped on 
each pass to the node $(w,\ell)$ 
($0\le w \le 2^{\ell}-1$, $0\le \ell \le n-1=2$). The internal state generated by 
passing through $\mathcal{H}_{(w,\ell)}$ is entangled with that of each walker by
the gate $\widetilde{\mathcal{E}}_{n-1}$.
}
\label{new-routing}
\end{figure}

First we construct the following routing scheme $\widetilde{\mathcal{F}}$:
\begin{equation}
\widetilde{\mathcal{F}}\colon 
|0\ket_A |0,0\ket_B |0\ket_C|0\ket_D\mapsto
\frac{1}{\sqrt{|\mathscr{A}|}}\sum_{a\in\mathscr{A}}|0\ket_A |a,n\ket_B
|0\ket_C|0\ket_D
\label{modified-routing}
\end{equation}
(cf. \eqref{routing}). Note that, in this routing scheme $\widetilde{\mathcal{F}}$,
the address information is {\it not} encoded in the address state, 
which actually remains $|0\ket_A$ during the routing. Instead, to deliver the
$n+m$ walkers to the memory cells at $\mathscr{A}$, the Hadamard-like gates
are appropriately placed in the first binary tree on the top sheet, where
the $0$th walker travels. Let $l_{(w,\ell)}$  (resp. $r_{(w,\ell)}$)
($0\le w \le 2^{\ell}-1$, $0\le \ell \le n-1$)
be the number of designated memory cells whose ancestor is the left child 
node $(2w,\ell+1)$ (resp. right child node $(2w+1,\ell+1)$) of $(w,\ell)$. 
See Fig.~\ref{new-routing} for a simple example.
Then, we define the Hadamard-like gate $\mathcal{H}_{(w,\ell)}$ acting
on the internal state of the $0$th walker that moves to the node $(w,\ell)$
from its parent node:
\begin{equation}
\mathcal{H}_{(w,l)}:=\frac{1}{\sqrt{l_{(w,\ell)}+r_{(w,\ell)}}}
\label{Hadamard}
\begin{pmatrix}
\sqrt{l_{(w,\ell)}} & \sqrt{r_{(w,\ell)}} \\
\sqrt{r_{(w,\ell)}} & -\sqrt{l_{(w,\ell)}}
\end{pmatrix}_{C_0},
\end{equation}
which is given by $e^{i\theta}R_y(\theta)R_z(\pi)$ for $\theta=
2\tan^{-1}(\sqrt{r_{(w,\ell)}/l_{(w,\ell)}})$
and is 
reduced to the standard Hadamard gate if
$l_{(w,\ell)}=r_{(w,\ell)}=1$ ($\theta=\pi/2$). Using this gate with 
$\mathcal{R}_{(w,\ell)}$, $\mathcal{X}_{(w,\ell)}$ defined
in \eqref{shift} and \eqref{encoder-def}, we can actually 
realize $\widetilde{\mathcal{F}}$:
\begin{align}
&\widetilde{\mathcal{F}}=\sum_{w=0}^{2^n-1}\mathcal{X}_{(w,n)}
\widetilde{\mathcal{F}}^{(n|n-1)}\cdots \widetilde{\mathcal{F}}^{(2|1)}
\widetilde{\mathcal{F}}^{(1|0)}, 
\quad
\widetilde{\mathcal{F}}^{(\ell+1|\ell)}:=
\sum_{w=0}^{2^\ell-1}\mathcal{R}_{(w,\ell)}\widetilde{\mathcal{E}}_{n-1}
\mathcal{H}_{(w,\ell)}\mathcal{X}_{(w,\ell)},
\end{align}
where $\widetilde{\mathcal{E}}_{\ell}$ is defined by slightly modifying
$\mathcal{E}_\ell$ (eq.~\eqref{encoder}) as
\begin{equation}
\widetilde{\mathcal{E}}_{\ell}:=
\mathcal{E}^{(p|p-1)}_{\ell}\cdots\mathcal{E}^{(2|1)}_{\ell}
\mathcal{E}^{(1|0)}_{\ell}.
\end{equation}
Namely, $\widetilde{\mathcal{E}}_{n-1}$ entangles the internal states of the $0$th
walker with those of the other walkers:
\begin{equation}
\widetilde{\mathcal{E}}_{n-1}\colon
\sum_c |c\ket_{C_0}\bigotimes_{j=1}^{n+m-1}|0\ket_{C_j}\mapsto
\sum_c \bigotimes_{j=0}^{n+m-1}|c\ket_{C_j} \quad (c\in\{0,1\}).
\end{equation}
In Fig.~\ref{new-routing}, we pictorially show an
example of the modified routing
scheme on the top sheet.
%
%%%%%%%%%%%%%%%%%%%%%%%%%%%%%%%%%%%%%%%%%%%%%%%%%%%%%%%%%
%%%%%%%%%%%%%%%%%%%%%%%%%%%%%%%%%%%%%%%%%%%%%%%%%%%%%%%%

To properly retrieve the walkers loaded with the data using output scheme
 $\mathcal{F}^{\dagger}$, 
the address information $|a\ket_A$  must be encoded in the positions of the 
quantum walkers. 
Note that once the data have been loaded, it is no longer possible 
to retrieve the walkers using $\widetilde{\mathcal{F}}^{\dagger}$.
The querying scheme $\widetilde{\mathcal{Q}}$ corresponding to \eqref{query} 
is modified to encode the data as well as the address of the cell where the 
data is stored:
\begin{equation}
\widetilde{\mathcal{Q}}\colon
\frac{1}{\sqrt{|\mathscr{A}|}}\sum_{a\in\mathscr{A}}|0\ket_A |a,n\ket_B |0\ket_C|0\ket_D
\mapsto
\frac{1}{\sqrt{|\mathscr{A}|}}\sum_{a\in\mathscr{A}}|a\ket_A |a,n\ket_B |0\ket_C|x^{(a)}\ket_D.
\end{equation}
Explicitly it reads
\begin{equation}
\widetilde{\mathcal{Q}}=\sum_{a\in\mathscr{A}}|a,n\ket\bra a,n|_B
\otimes
\bigotimes_{i=0}^{n-1} \left(X_{A_i}\right)^{a_i}
\otimes
\bigotimes_{i=0}^{m-1} \left(X_{D_i}\right)^{x_i^{(a)}},
\end{equation}
which is accomplished
by exchanging the specified paths in the address and data register, 
or alternatively by placing the Pauli-X gates, as explained in the
previous section.

Finally applying the output scheme $\mathcal{F}^{\dagger}$ defined in \eqref{output}, 
we obtain the desired qRAM architecture \eqref{qRAM3} in the form
\begin{equation}
\widetilde{\mathrm{qRAM}}=\mathcal{F}^{\dagger}\widetilde{\mathcal{Q}}
\widetilde{\mathcal{F}}.
\end{equation}

%%%%%%%%%%%%%%%%%%%%%%%%%%%%%%%%%%%%%%%%%%%%%%%%%%%%%%%%%%%%
\section{Summary and discussion}
%%%%%%%%%%%%%%%%%%%%%%%%%%%%%%%%%%%%%%%%%%%%%%%%%%%%%%%%%%%%
A qRAM \eqref{qRAM2} or \eqref{qRAM3} has been physically realized by 
combinations of several elementary 
quantum devices, including the roundabout gate \eqref{fig-RA} developed in 
the first paper \cite{asaka2021two} in this series. $2^n$ $m$-qubit information can 
be retrieved in superposition by simply passing the $n+m$ quantum walkers
through the perfect binary trees, as schematically shown in Fig.~\ref{overview}. 
The advantages of the present qRAM architecture compared to the original
bucket-brigade qRAM are
summarized as follows: (i) The procedure is completely parallelized without using
any ancilla qubit. The $2^n$ $m$-qubit information can be retrieved after
$O(n\log(n+m))$ steps. The qubit resources and the quantum gates
required for the processing are $O(n+m)$ and $O(2^n (n+m))$.
(ii) The walkers do not entangle with any device on the
binary trees, which promises to reduce  the cost of maintaining quantum coherence.
(iii) Our qRAM architecture is free from any time-dependent control. In other words, 
information in quantum superposition can be composed by just passing the walkers through
the binary trees. On the other hand, the trade-off for these advantages is that 
it requires more space and quantum gates.

\begin{figure}
\centering
\includegraphics[width=0.9\textwidth]{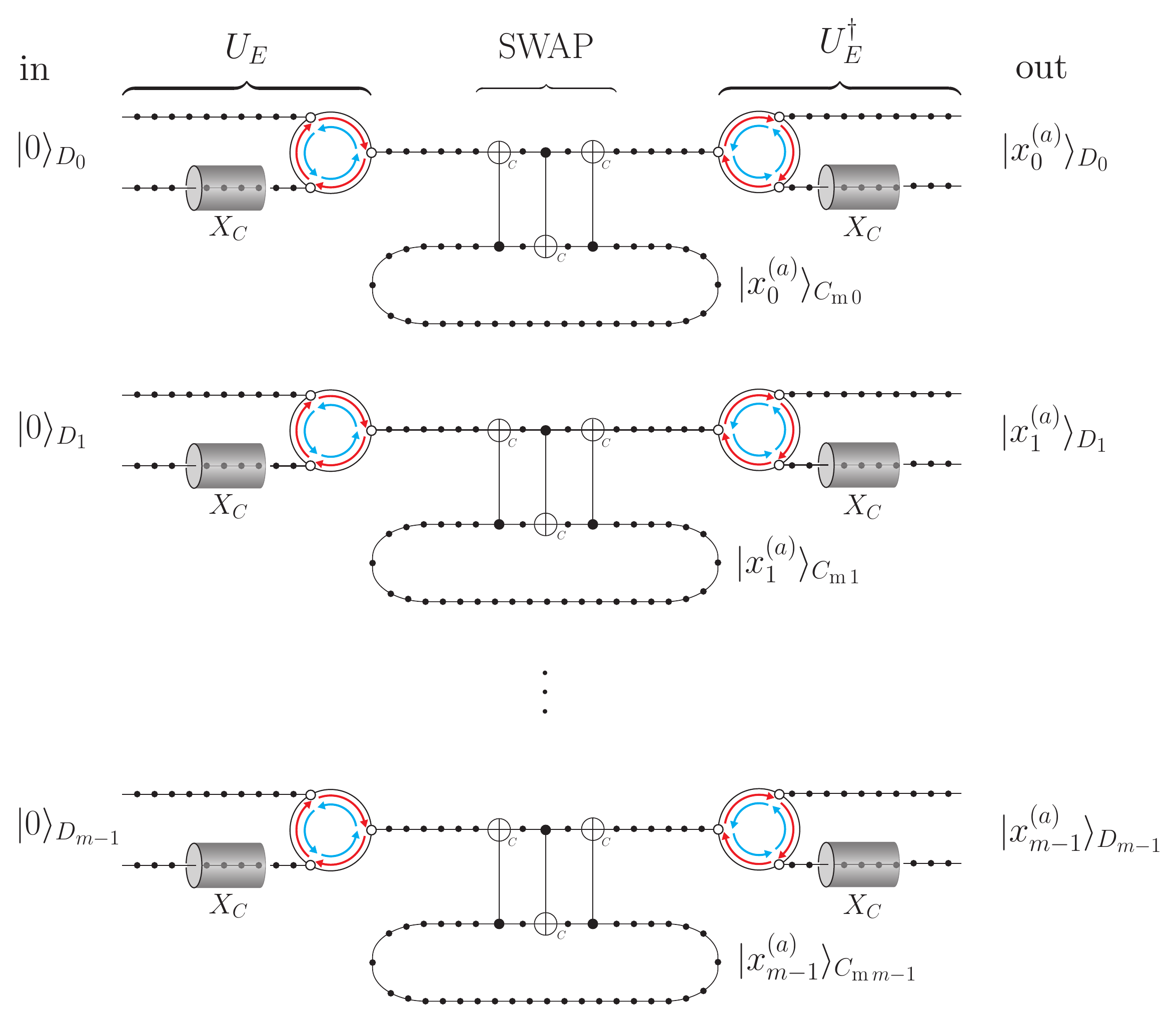}
\caption{A physical realization of the process \eqref{q-info}.
Quantum information is represented as the internal states
of quantum walkers, which can be stored in the memory 
cell using a \textit{quantum memory} developed  in the first paper
\cite{asaka2021two}.
The information stored in the memory cell is transfered by
the encoder/decoder $U_\mathrm{E}$/$U^{\dagger}_\mathrm{E}$
and the swap gate consisting of the CNOT gates
defined in \eqref{cnotc2}.
}
\label{quantum}
\end{figure}

Finally, let us discuss how to generalize the present architecture to be 
able to process quantum information (i.e. information stored in the cell
consists of a superposition of states). In the querying scheme described in
Sec.~3, $m$-bit classical information stored in a memory cell can be 
directly copied to 
an $m$-qubit state of the data register (see \eqref{query} and 
\eqref{querying}), which can be achieved by exchanging the appropriate paths 
in the data register or by placing the Pauli-X gates. For the quantum case,
however, quantum infromation can not be replicated due to the no-cloning 
theorem \cite{wootters1982single,dieks1982communication}. Instead, by a
swap gate, the quantum information can be transferd to the state in 
the data register. 
In our architecture,  $m$-qubit quantum information stored in the 
cell at address $a$ can be represented as the internal states of $m$ 
quantum walkers, each moving in a circle, as shown in Fig.~\ref{quantum}.
(This may be realized by a \textit{quantum memory} developed 
in Sec.~6 in the first paper \cite{asaka2021two}).
Explicitly, it is expressed as $\sum_x |x^{(a)}\ket_{C_{\mathrm{m}}}$,
where the index $C_\mathrm{m}$ stands for the colors of the 
quantum walkers in the memory cell.  
As shown in Fig.~\ref{quantum}, combining an encoder
$U_\mathrm{E}$ (see eq. (4.2) in the first paper \cite{asaka2021two} for a
precise definition),
which encodes the position of the quantum walker into the internal 
state of the walker, and the swap gate consisting of the three CNOT gates
$\mathrm{CX}_{C_jC_k}$ in \eqref{cnotc2}, we can correctly transfer
the quantum information in the memory cell to the data register:
\begin{equation}
\sum_x |x^{(a)}\ket_{C_\mathrm{m}} \mapsto \sum_x |x^{(a)}\ket_{D}.
\label{q-info}
\end{equation}
%
%%%%%%%%%%%%%%%%%%%%%%%%%%%%%%%%%%%%%%%%%%%%%%%%%%%%%%%%%
%%%%%%%%%%%%%%%%%%%%%%%%%%%%%%%%%%%%%%%%%%%%%%%%%%%%%%%%

Combining the model of universal quantum computation achieved in the first paper
\cite{asaka2021two}  with the current qRAM architecture is expected to 
enable efficient processing of quantum information such as the quantum phase 
estimation \cite{shor1994algorithms}, Grover's algorithm for searching 
unsorted databases \cite{grover1996fast}, and the quantum 
version of fast Fourier transform \cite{asaka2020quantum}.
%%%%%%%%%%%%%%%%%%%%%%%%%%%%%%%%%%%%%%%%%%%%%%%%%%%%%%%%%%%%
\section*{Acknowledgment}
%%%%%%%%%%%%%%%%%%%%%%%%%%%%%%%%%%%%%%%%%%%%%%%%%%%%%%%%%%%%
%
The present work was partially supported by Grant-in-Aid for Scientific
Research (C) No. 20K03793 from the Japan Society for the 
Promotion of Science.

\bibliography{BibFile}

\end{document}